\documentclass[twocolumn, tighten, trackchanges, resetfootnote]{aastex7}

\usepackage{amsmath}

\makeatletter
\providecommand{\onecolumngrid}{\onecolumn}
\providecommand{\twocolumngrid}{\twocolumn}
\makeatother

\shorttitle{Wide Binaries of extreme separations}
\shortauthors{Yoon et al.}

\received{July 17, 2025}
\revised{\today}
\accepted{}

\submitjournal{ApJ}

\begin{document}

\title{Probing the Nature of Gravity in the Low-acceleration Limit: \\Wide Binaries of Extreme Separations with Perspective Effects}

\author[orcid=0000-0002-0096-4702, gname=Youngsub, sname=Yoon]{Youngsub Yoon}\thanks{youngsuby@gmail.com}
\affiliation{Department of Physics and Astronomy, Sejong University, 209 Neungdong-ro Gwangjin-gu, Seoul 05006, Republic of Korea}
\email{youngsuby@gmail.com}
\author[orcid=0000-0001-9962-1816, gname=Yong, sname=Tian]{Yong Tian}
\email{yongtian@phy.ncu.edu.tw}\thanks{yongtian@phy.ncu.edu.tw}
\affiliation{Department of Physics and Astronomy, Sejong University, 209 Neungdong-ro Gwangjin-gu, Seoul 05006, Republic of Korea}
\affiliation{Department of Physics, National Central University, Taoyuan 320317, Taiwan}
\author[orcid=0000-0002-6016-2736, gname=Kyu-Hyun, sname=Chae]{Kyu-Hyun Chae}
\email{chae@sejong.ac.kr}\thanks{corresponding author: chae@sejong.ac.kr \\ kyuhyunchae@gmail.com}
\affiliation{Department of Physics and Astronomy, Sejong University, 209 Neungdong-ro Gwangjin-gu, Seoul 05006, Republic of Korea}

\begin{abstract}

Recent statistical analyses of wide binaries have revealed a boost in gravitational acceleration with respect to the prediction by Newtonian gravity at low internal accelerations $\lesssim 10^{-9}$ m\,s$^{-2}$. This phenomenon is important because it does not permit the dark matter interpretation, unlike galaxy rotation curves.
We extend previous analyses by increasing the maximum sky-projected separation from 30 to 50 kilo astronomical units (kau). We show that the so-called ``perspective effects'' are not negligible at this extended separation and, thus, incorporate it in our analysis. With wide binaries selected with very stringent criteria, we find that the gravitational acceleration boost factor, $\gamma_g \equiv g_{\rm obs}/g_{\mathrm N}$, is $1.61^{+0.37}_{-0.29}$ (from $\delta_{\rm obs-newt}\equiv (\log_{10}\gamma_g)/\sqrt{2}=0.147\pm0.062$) at Newtonian accelerations $g_{\mathrm N} = 10^{-11.0}$ m\,s$^{-2}$, corresponding to separations of tens of kau for solar-mass binaries. At Newtonian accelerations $g_{\mathrm N} = 10^{-10.3}$ m\,s$^{-2}$, we find $\gamma_g=1.26^{+0.12}_{-0.10}$ ($\delta_{\rm obs-newt}=0.072\pm0.027$). For all binaries with $g_{\rm N}\lesssim10^{-10}$ m\,$s^{-2}$ from our sample, we find $\gamma_g=1.32^{+0.12}_{-0.11}$ ($\delta_{\rm obs-newt}=0.085\pm0.027$). These results are consistent with the generic prediction of MOND-type modified gravity, although the current data are not sufficient to pin down the low-acceleration limiting behavior. 
Finally, we emphasize that the observed deviation from Newtonian gravity cannot be explained by the perspective effects or any separation-dependent eccentricity variation that we have taken into account.
\end{abstract}

\keywords{\uat{Binary stars}{154} --- \uat{Gravitation}{661} --- \uat{Modified Newtonian dynamics}{1069} --- \uat{Non-standard theories of gravity}{1118} --- \uat{Wide binary stars}{1801}}

\section{Introduction}\label{sec:intro}

In 1983, Milgrom proposed Modified Newtonian Dynamics \citep[MOND, ][]{Milgrom:1983ca} as an alternative path to the dark matter problem. For MOND reviews, the reader is referred to \cite{Sanders:2002}, \cite{Famaey:2012}, and \cite{Banik:2022}. MOND posits an acceleration scale $a_0$ ($\approx 1.2\times 10^{-10}$ m\,s$^{-2}$) as a new physical constant that marks the scale where standard gravity breaks down. This basic tenet of MOND can be most directly tested by wide binaries because any hypothetical dark matter cannot play any role in their internal dynamics \citep{Hernandez:2011uf}.
When the two stars of a binary are separated by more than several kau, their mutual internal acceleration gets $\la a_0$ and, according to MOND, the norm of the relative velocity between the pair must be greater than the Newtonian expectation.
On the other hand, dark matter can hardly change the relative velocity, given that its hypothetical total mass within a wide binary orbit is smaller than the wide binary mass by more than four orders of magnitude. The dark matter density near the Sun is estimated to be only $10^{-9} M_{\odot}$/kau$^3$ \citep{Read}. Therefore, the mass of dark matter inside a few kau sphere is less than $10^{-4} M_{\odot}$. Thus, if wide binary orbit anomalies are experimentally proven, it means a falsification of the assumption of standard gravity behind dark matter.

By analyzing wide binaries from the Gaia data release (DR) 3 database \citep{Gaia:2023}, Chae \citep{Chae2023a, Chae2023b, Chae2024, Chae2025} and Hernandez et al. \citep{Hernandez:2023qfj, Hernandez:2023upr, HernandezKroupa} independently concluded that gravitational acceleration is boosted by a factor of $\gamma_g\approx 1.4-1.5$ at Newton-predicted gravitational acceleration $g_{\rm N}\lesssim 10^{-9.5}$ m\,s${^{-2}}$ in agreement with the generic predictions of Lagrangian theories of nonrelativistic MOND gravity such as AQUAL \citep{Bekenstein:1984tv} and QUMOND \citep{Milgrom:2009ee} and relativistic MOND gravity \citep[see e.g.,][]{Mistele:2024}. These theories break the strong equivalence principle (while keeping Galilei's universality of free fall), necessitating the external field effect (EFE). According to such theories, wide binaries in the solar neighborhood are subject to an EFE from the rather strong external field $\approx 1.8 a_0$ from the Milky Way and are consequently expected to obey a rather mild gravity boost of $\approx 1.4$.  

However, based on the currently available data, there are two major observational limitations in using wide binaries to check whether they deviate from Newtonian predictions. First, Newtonian gravity predicts an elliptical orbit for a bound binary system, and a significant segment of the orbit is ideally needed to determine the orbital shape, but the DR3 time span is 34 months, while the orbital periods of typical wide binaries are on the order of hundreds of thousands of years. Second, even from the ``snapshot'' observations of wide binaries, not all of the six components of the relative displacement and relative velocity between the pair are accurately measured. In general, the line-of-sight separation is essentially unknown or ill-determined (as the distance errors are usually much larger than the relatively small orbital sizes although they are ``wide'' binaries), and radial velocities of the stars, in general, have large uncertainties (or are not available at all) from the Gaia DR3. Therefore, most studies used statistical methods based only on the 2D relative position and the 2D relative velocity projected on the sky plane, which are in general accurately measured from Gaia observations (see Table 3 of \cite{Chae2025} for a summary of statistical methods). 

Recently, new approaches beyond statistical methods relying only on 2D quantities have been proposed or considered. \cite{Chae2025} considered a sample of 312 wide binaries selected from Gaia DR3 that have relatively accurate radial velocities, while \cite{Saglia2025} considered a smaller sample of 32 binaries (including only several with internal acceleration $\la a_0$) that have the HARPS radial velocities whose precision is typically tens of m\,s$^{-1}$ (an order of magnitude more precise than the currently available Gaia radial velocities). \cite{Chae2025} also employed a novel Bayesian method to derive a probability distribution of the gravity boost factor. As regards these new developments, \cite{Saglia2025} is limited by the small sample size while \cite{Chae2025} is limited by the relatively larger uncertainties of radial velocities, and both are limited by insufficiently precise distances. However, these approaches will become more and more fruitful as more and more sufficiently precise radial velocities become available in the future.

Regardless of methods to probe gravity, wide binaries with extremely large separations have not yet been analyzed properly, although they are potentially of great importance because they can be used to probe the low-acceleration limiting behavior of gravity. Use of extremely wide binaries involves additional difficulties. First, it is more difficult to find truly isolated and gravitationally-bound wide binaries as the separation gets larger. Thus, extreme care is needed to select those binaries. Second, the relative velocity between the pair gets smaller at larger separation, and thus greater measurement precision is required. Finally, the so-called ``perspective effects'' \citep{Shaya:2010hs, El-Badry2019,Hernandez:2023qfj,Hernandez:2023upr} are important and have to be taken into account. When measuring the relative velocity, measured velocities can simply be subtracted from each other for binaries with small separations, as their relative positions can be regarded as being placed on a flat surface due to the negligible curvature of the celestial sphere. However, this is not the case for those with large separations, especially if the barycenter of the binary system has a large velocity relative to the Sun.

In this study, we select wide binaries with separations up to 50 kau to probe gravity in the low-acceleration limit. The limit of 50 kau is significantly larger than previous limits used in quantitative statistical analyses of wide binaries. In previous studies, the nominal maximum limit was 30 kau (e.g., \citealt{Chae2023a}), but most wide binaries had separations smaller than about 20 kau. As will be shown, the perspective effects are not negligible for our sample and are therefore taken into account.

The organization of this paper is as follows. 
In Section \ref{sec:selectioncriteria}, we describe our selection criteria for binaries. In Section \ref{sec:perspective}, we describe how we calculate the perspective effects. In Section \ref{sec:scalingrelation}, we describe some scaling relations in our sample. In Section \ref{sec:MonteCarlo}, we present our Monte Carlo (MC) simulation results. Discussions and conclusions are given in Section \ref{sec:conclusion}. The newly defined wide binary samples are available on Zenodo under an open-source Creative Commons Attribution license: \dataset[10.5281/zenodo.16934778]{https://doi.org/10.5281/zenodo.16934778}. The codes used for statistical analyses can be found at \cite{Chae2023Zenodo}.

\section{Wide Binaries Sample}\label{sec:selectioncriteria}

Our analysis is based on the comprehensive catalog of wide binaries constructed by \citet{El-badrysample} from the Gaia DR3 database. This catalog is based on astrometric and photometric quality cuts, and includes pairs with projected separations from a few au up to 1 pc ($2 \times 10^5$ au) and distances within 1 kpc from the Sun. Importantly, \citet{El-badrysample} empirically estimated the probability of chance alignment $R$ for each pair, providing a reasonable means to isolate gravitationally bound binaries. Their catalog contains approximately $1.3$ million binaries with $>90\%$ probability of being gravitationally bound, and over $1.1$ million with a $>99\%$ probability, making it the largest and most reliable sample of wide binaries currently available. Bound binaries in this context are defined as pairs whose kinematics and parallaxes are consistent, within uncertainties, with being gravitationally bound.

To ensure high purity and minimize contamination from chance alignments and problematic measurements while keeping a non-negligible number of binaries with extreme separations, we apply the following selection criteria in defining our working sample of wide binaries. First, we select only pairs with $R < 0.1$.  
Second, we require that both components are within $300$ pc, which focuses our sample on nearby, well-measured systems. These two criteria reduce the number of binaries to 264,483. Third, we restrict the projected separation $s$ to the range $0.2~{\rm kau} < s < 50~{\rm kau}$ to probe the low-acceleration regime up to $\approx 10^{-11}$ m\,s$^{-2}$ while having a broad dynamic range from the deep Newtonian regime. 
Finally, we require that both stars in a binary have exceptionally precise proper motions (PMs) and parallaxes with fractional measurement errors less than 0.5\%. This ensures that every star, regardless of their membership, satisfies the same data qualities and the same probability of contamination from, e.g., unseen faint close companion stars or massive Jovian planets.
After all these selection criteria and quality cuts, our final sample consists of 26,970 wide binaries with high probability of being bound and exceptionally precise astrometric data, suitable for testing gravity theories at separations up to $50~{\rm kau}$.

The \cite{El-badrysample} sample selection process removed observationally resolved multiples with their own criteria, meaning that our above-selected sample is also likely to be largely free from resolved multiples.
However, to further purify the sample, we implement a stricter isolation criterion based on the local environment of each candidate binary. For each system, we first counted the number of neighboring stars that satisfy the following two conditions:
\begin{equation}\label{eq:isolation_sep}
    s'<\sqrt{20} s
\end{equation}
\begin{equation}\label{eq:conditionddA}
    |d-d_A|<2\sqrt{\sigma_{d_A}^2+\sigma_{d_B}^2} +4 s'
\end{equation}
where $s'$ is the sky-projected separation from a given star to the primary (A), and $d$ is the distance to the star. In practice, we selected only those binaries for which exactly one star—namely the designated companion B—satisfies both criteria, in addition to the primary A. If no additional star satisfies these criteria, the pair (A and B) could be a chance alignment (so-called ``line-of-sight'' contamination) rather than a true bound binary. Conversely, if a third star meets these conditions, it could dynamically perturb the system and render Keplerian orbital calculations unreliable. The factor $\sqrt{20}$ in Equation~(\ref{eq:isolation_sep}) is chosen such that any perturbation to the gravitational force is less than about 5\%.

\begin{figure*}[!htb]
    \centering
    \includegraphics[width=1.9\columnwidth]{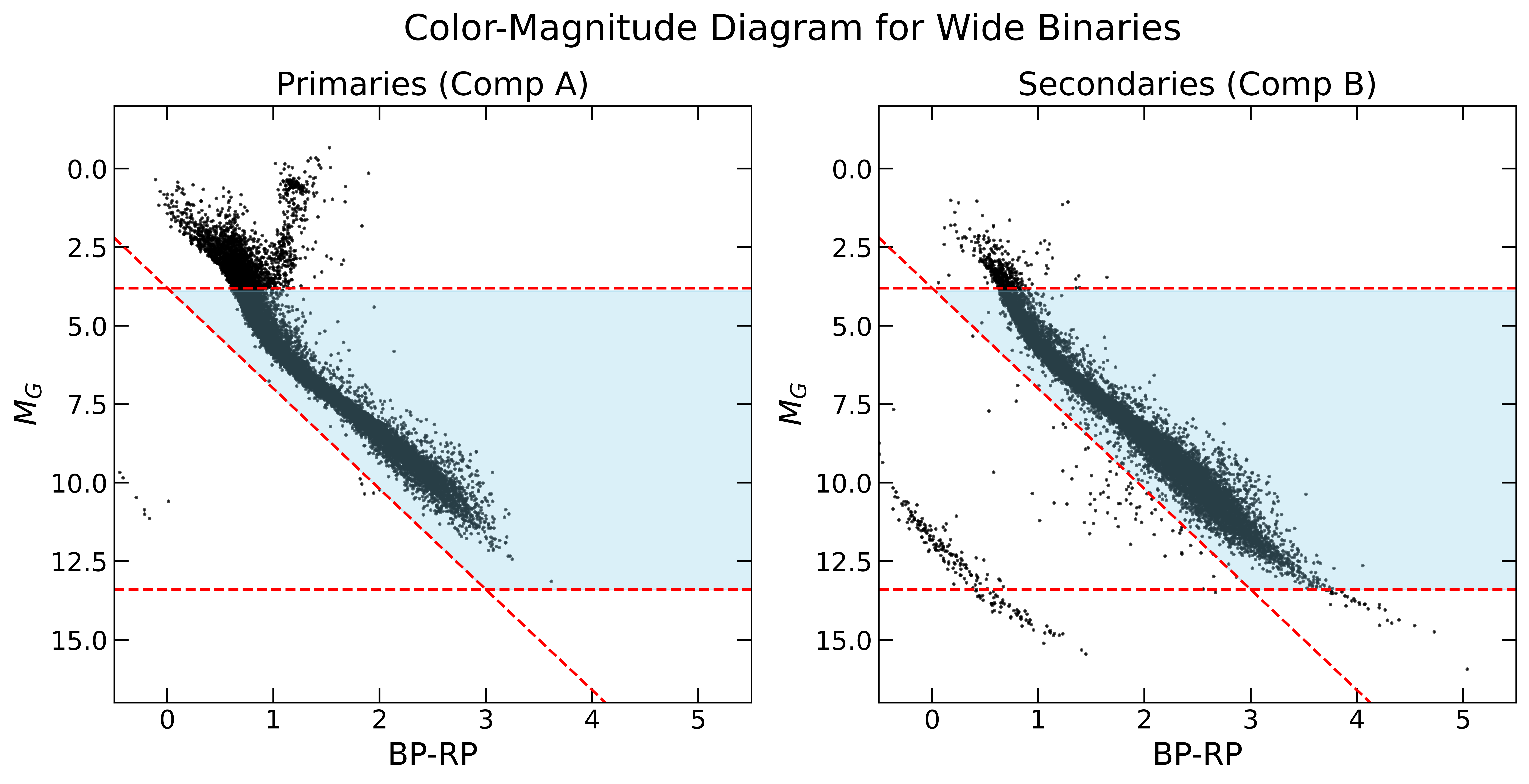}
    \caption{
    Color--magnitude diagrams for the primary (component~A, left panel) and secondary (component~B, right panel) stars in our wide binary sample. Each point represents a star, plotted in absolute $G$-band magnitude ($M_G$) versus \emph{Gaia} color ($BP-RP$). The shaded blue region denotes the selection window applied to ensure that both components are main-sequence stars and to exclude evolved objects and outliers. The dashed red lines indicate the magnitude and color boundaries: $3.8 < M_G < 13.4$ and $M_G - 3.2(BP-RP) < 3.8$. Only binaries in which both components fall within the shaded region are retained for subsequent analysis.
    }
    \label{fig:HRdiagram}
\end{figure*}

The above added selection criteria of Equations~(\ref{eq:isolation_sep}) and (\ref{eq:conditionddA}) have effectively removed any remaining chance alignments and resolved multiples from the \cite{El-badrysample} sample. Also, the requirement of high precision for PMs and parallaxes of individual stars has significantly reduced the probability that a star has unresolved kinematic contaminants such as low mass companion stars or massive Jovian planets. Nevertheless, there will certainly remain individual exceptions with unresolved kinematic contaminants. To help further remove those contaminants and ensure other data qualities such as stellar masses, we apply several additional criteria, following \citet{Chae2023a, Chae2023b}:

\begin{itemize}
\item Both components must have Gaia \texttt{ruwe} values less than 1.2.
\item The following distance consistency condition is required:
\begin{equation}
    |d_A - d_B| < \sqrt{4(\sigma_{d_A}^2 + \sigma_{d_B}^2) + (6s)^2}.
\end{equation}
\item At least one component must have a measured radial velocity from the Gaia DR3 database.
\item If both radial velocities are available, they must satisfy the following consistency condition:
\begin{equation}
    |v_{r,A}-v_{r,B}|<\sqrt{4(\sigma_{v_{r,A}}^2+\sigma_{v_{r,B}}^2)+(\Delta v_{r,\mathrm{orbit}}^{\mathrm{max}})^2},
    \label{eq:velbound}
\end{equation}
where 
\begin{equation}
    \Delta v_{r,\mathrm{orbit}}^{\mathrm{max}} = 0.9419 ~\mathrm{km} ~\mathrm{s^{-1}} \sqrt{\frac{M_{\mathrm{tot}}}{s}}\times 1.3\times 1.3
\end{equation}
gives an upper limit for the physical relative radial velocity allowed for a binary with projected separation $s$. The first factor of 1.3 accounts for geometric effects from orbital inclination and phase, while the second accounts for a possible boost in velocity due to gravitational anomalies.
\item Both stars must have absolute $G$-band absolute magnitudes between 3.8 and 13.4, and satisfy the main-sequence color-magnitude criterion: $M_G - 3.2 (BP - RP) < 3.8$ \citep{Penoyre}. These criteria, illustrated by the shaded regions in Figure~\ref{fig:HRdiagram}, ensure a clean main-sequence selection.
\item As in \citet{Chae2023a}, we account for dust extinction using a 3D dust map available only for declinations above $-28^\circ$; therefore, binaries located below this declination are excluded, removing about 30\% of the data.
\item The impact of radial velocity uncertainty on the relative PM (when considering the perspective effects) must be less than 5\%. Details of this criterion and its impact are discussed in Section~\ref{sec:perspective}.
\end{itemize}

In the above selection, we adopt that both stars have fractional PM uncertainties less than 0.5\%. To assess the impact of astrometric precision, we consider an alternative criterion for sample selection: the error of the relative (sky-projected) 2D velocity is smaller than 50~m\,s$^{-1}$. These two criteria produced largely overlapping samples: the first yielded 8,095 binaries, the second 8,806, with 8,068 systems in common. For systems with $s > 30$ kau, the first criterion selected 33 binaries and the second 31, with 29 overlapping. This is because most binaries already satisfy the second criterion. See the inset panel of Figure~\ref{fig:sigmavp}.

\section{Perspective effects}\label{sec:perspective}
Gaia recorded the PM of each star.
A PM in the projected 2D direction can be represented as follows.
\begin{equation}
\vec \mu=(\mu_\alpha, \mu_\delta)\equiv\left(\frac{\vec v\cdot \hat \alpha}{d}, \frac{\vec v \cdot \hat \delta}{d}\right),   
\end{equation}
where $\alpha$ denotes the right ascension and $\delta$ denotes the declination. Then, the relative PM is given by
\begin{equation}
    \Delta \vec \mu \equiv \vec \mu_A-\vec \mu_B.
\end{equation}
Multiplying this by the distance, we can obtain the relative velocity between the two components of the binary. However, in obtaining the above expression, the key assumption was that the directions of the coordinates $\hat \alpha$ and $\hat \delta$ are the same for both components of the binary, A and B. This assumption is valid if the angular separation between the two stars is negligible. For a large separation, which we consider in this study, this is no longer the case. The consideration of such a large separation in the calculation of the relative velocity is known as the perspective effects \citep{Shaya:2010hs}. In this work, we consider the perspective effects and confirm that they are not negligible for the largest separations. In the following, we outline our approach to account for the perspective effects. In particular, we will see that we do not need to consider the galactic coordinates as in \cite{Shaya:2010hs}. Instead, we can directly work with the celestial coordinates.

\begin{figure*}[!htb]
    \centering
    \includegraphics[width=2.1\columnwidth]{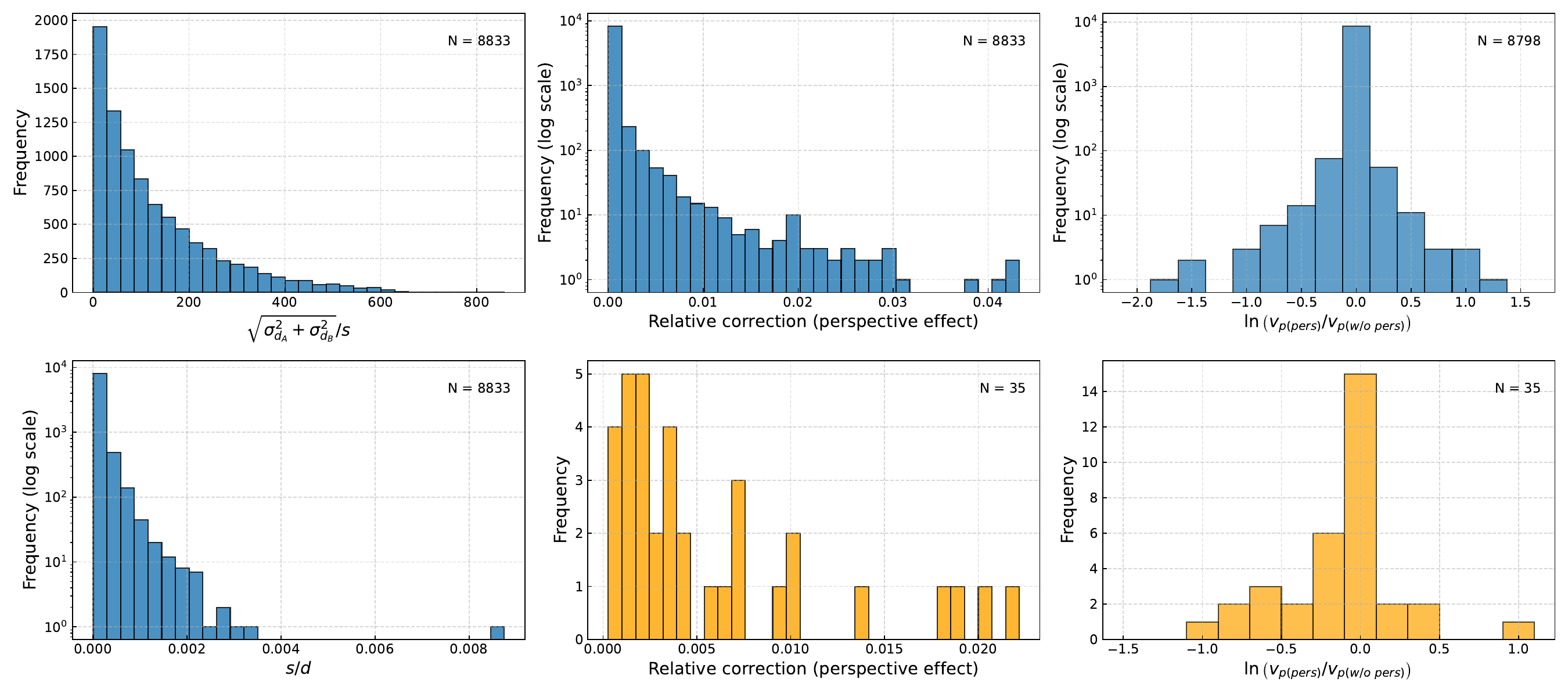}
    \caption{
    Summary of key error terms and the perspective effects for the wide binary sample, with each panel illustrating a different aspect of the analysis. 
    \textbf{Top-left:} Assessment of line-of-sight distance uncertainties in the context of the perspective effects. The distribution of $\sigma_{\Delta d}/s=\sqrt{\sigma_{d_A}^2+\sigma_{d_B}^2}/s$ is shown; its large value (typically $\mathcal O(100)$) makes it impossible to correctly estimate the $\Delta d/d$ term in the perspective effects because the true value of $\Delta d$ is on the order of $s$. Our nominal choice is to assume that the $\Delta d/d$ term is zero.  
    \textbf{Bottom-left:} Distribution of $s/d$, representing the expected distribution of $\sqrt{2}|\Delta d|/d$. 
    \textbf{Top-center:} Distribution of the relative error in the relative PM due to radial velocity uncertainty (perspective effects). The fractional error is generally small for most wide binaries.
    \textbf{Bottom-center:} Same as the top-center panel, but only for systems with sky-projected separation $s > 30$ kau. In this regime, the impact of perspective effects becomes most significant.
    \textbf{Top-right:} Distribution of $\ln \left( v_{p(\mathrm{pers})}/v_{p(\mathrm{w/o~pers})} \right)$ for binaries with $s < 30~\mathrm{kau}$, quantifying the typical change in projected relative velocity due to the perspective effects at smaller separations.
    \textbf{Bottom-right:} Same as the top-right panel, but for binaries with $s > 30~\mathrm{kau}$. The perspective effects become significant only at the largest separations.
    }
    \label{fig:fig2}
\end{figure*}

Using the following notations for the angular separation and the distance separation
\begin{equation}
    \Delta \alpha \equiv \alpha_A-\alpha_B,\quad \Delta \delta \equiv \delta_A-\delta_B,\quad \Delta d \equiv d_A-d_B,
\end{equation}
The consideration of the perspective effects changes the relative PM in the first order in the separation as follows:
\begin{equation}
    \Delta \vec \mu \rightarrow \Delta \vec \mu + (\Delta \mu_{\alpha(\mathrm{pers})},\Delta \mu_{\delta(\mathrm{pers})}),
\end{equation}
\begin{equation}
    \Delta \mu_{\alpha(\mathrm{pers})} = \mu_\alpha \frac{\Delta d}{d} +\left(\frac{v_r}{d}\cos\delta-\mu_\delta\sin\delta\right)\Delta\alpha,
    \label{eq:mupersalpha}
\end{equation}
\begin{equation}
    \Delta \mu_{\delta(\mathrm{pers})} = \mu_\delta \frac{\Delta d}{d} +\mu_\alpha\sin\delta~\Delta\alpha +\frac{v_r}{d}\Delta\delta,
    \label{eq:mupersdelta}
\end{equation}
where $\mu_\alpha$, $\mu_\delta$ are the system PM and $v_r\equiv\vec v \cdot \hat r$, the system radial velocity. 
Therefore, apart from the term proportional to $\Delta d/d$, i.e., the first terms in Equations~(\ref{eq:mupersalpha}) and~(\ref{eq:mupersdelta}), the perspective effects are important when the velocity of the system, i.e. binary as a whole, multiplied by the angular separation (in radian) is not negligible compared to the relative velocity between the two stars in the binary.

The radial velocity is also subject to the perspective effects. In this study, we do not directly use the radial velocity, except to estimate the perspective effects. Nevertheless, for completeness, we present the formula for the change in the relative radial velocity due to the perspective effects. For $\Delta v_r\equiv v_{rA}-v_{rB}$, we have
\begin{equation}
    \Delta v_r\rightarrow\Delta v_r -d(\mu_\alpha \cos\delta~ \Delta\alpha +\mu_\delta \Delta \delta).
    \label{eq:pers_rad}
\end{equation}
We note that the analytically calculated perspective effects given by Equations~(\ref{eq:mupersalpha}), (\ref{eq:mupersdelta}), and (\ref{eq:pers_rad}) agree well with direct numerical calculations with the algorithm described in Appendix~B of \cite{Chae2025}. 

These equations work excellently even for the Proxima - $\alpha$ Cen AB system, the nearest wide binary from the Sun. Our analytic results for this system are expected to show the largest deviation from the direct numerical calculations by \cite{Kervella}. However, the difference is only about 0.07 m/s, totally negligible compared to the real relative velocity of 273 m/s, and far less than 4 m/s, the error of proper motion component velocity, and 32 m/s, the error of radial velocity components. Roughly speaking, the PM motion is given by $\Delta \mu$ in the lowest order, and the perspective effect formulas above give the corrections on the order of $\mathcal O(\mu \Delta \theta)$ where $\theta$ denotes the angular separation $\alpha$ and $\delta$. Skipping $\mathcal O(\Delta \mu \Delta \theta)$ and $\mathcal O(\mu \Delta \theta^2)$, the next corrections come at the order of $\mathcal O(\Delta \mu \Delta \theta^2), \mathcal O(\mu \Delta \theta^3)$, which are completely negligible, as we just explained.

In all these formulas for the perspective effects, every symbol without $\Delta$ denotes the system value, the value of the binary as a whole. For example,
\begin{equation}
    \alpha = (\alpha_A+\alpha_B)/2, \quad \delta=(\delta_A+\delta_B)/2,
\end{equation}
\begin{equation}
    \mu_\alpha = (\mu_{\alpha A}+\mu_{\alpha B})/2, \quad \mu_\delta=(\mu_{\delta A}+\mu_{\delta B})/2.
\end{equation}
In other words, they are the average values of each component. However, with respect to the system distance, it is more accurate to obtain it by its weighted average considering the errors of the distances to both components, as both distances have much larger errors than their distance difference, the real value of $\Delta d~(=d_A({\rm real})-d_B({\rm real}))$. In other words, for the observed values of $d_A$ and $d_B$, we use
\begin{equation}
    d= \frac{d_A/\sigma_{d_A}^2+d_B/\sigma_{d_B}^2}{1/\sigma_{d_A}^2 +1/\sigma_{d_B}^2}.
\end{equation}
Similarly, as the errors of the radial velocities are much larger than $\Delta v_r~(=v_{rA}({\rm real})-v_{rB}({\rm real}))$, for the observed values of $v_{rA}$ and $v_{rB}$, we have
\begin{equation}
    v_r= \frac{v_{rA}/\sigma_{v_{rA}}^2+v_{rB}/\sigma_{v_{rB}}^2}{1/\sigma_{v_{rA}}^2 +1/\sigma_{v_{rB}}^2}.
\end{equation}
In cases where the radial velocity of only one of the two stars is available, we assume that the system radial velocity is given by the radial velocity of the available one.

In considering the perspective effects, our nominal choice will be to ignore the term proportional to $\Delta d/d$ for the following reasons. The error in the observed value of $\Delta d$ given by
\begin{equation}
    \sigma_{\Delta d}=\sqrt{\sigma_{d_A}^2+\sigma_{d_B}^2}\,,
\end{equation}
is about two orders of magnitude greater than the true value of $\Delta d$, which should be in the order of $s$, the sky-projected separation. See the top-left panel of Figure~\ref{fig:fig2} for a histogram of $\sigma_{\Delta d}/s$. Thus, considering the first term would induce errors that would be much larger than those without it. Lacking accurate values of $\Delta d$, we could estimate the effect of the term proportional to $\Delta d/d$ using realistic MC generated values of $\Delta d$, but it is not warranted in a statistical sense because $\Delta d/d$ can be positive or negative with equal probabilities, see Appendix~B of \cite{Chae2025}.

\begin{figure}[!htb]
    \centering
    \includegraphics[width=1.0\columnwidth]{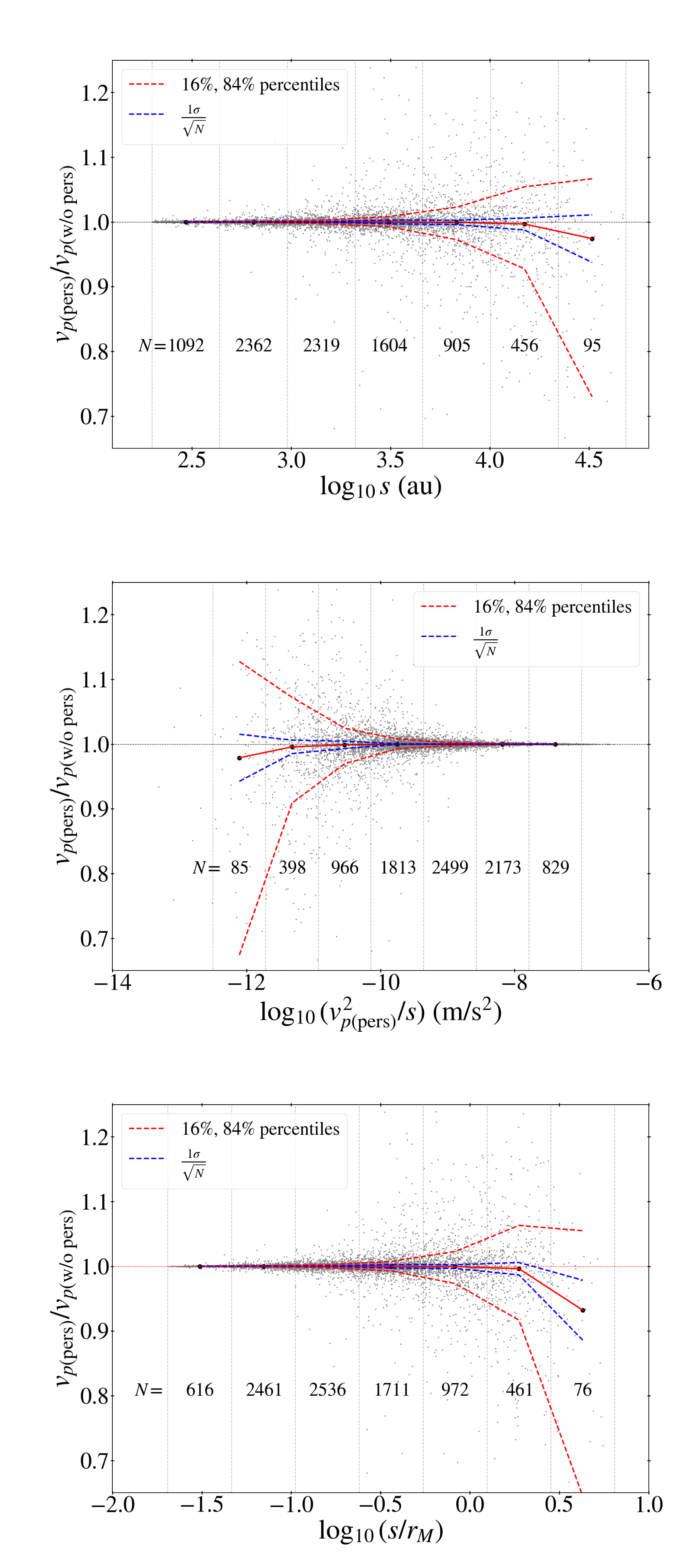}
    \caption{perspective effects as a function of three variables: $s$, $v^2_{\rm pers}/s$ and $s/r_M$. The dots represent $v_{\rm pers}/v$ and the solid red lines denote the median for each bin. For all three graphs, the perspective effects are important only for the last bin.
     }
    \label{fig:perspec}
\end{figure}

In contrast to $d$, the error of $v_r$ is small enough for it to be considered in the perspective effects. Before imposing the last condition for our sample selection, we had 8,847 binaries. Among those, only 14 binaries would have more than 5\% of the relative error of the relative PM due to the error of $v_r$, when the perspective effects are considered. We remove these 14 data points. This is the last condition for our sample selection. This criterion does not significantly harm our analysis of binaries with a large separation since the number of data points with $s > 30~{\rm kau}$ is reduced from 37 to 35. In the middle column of Figure~\ref{fig:fig2}, the distribution of the relative error that the relative PM would have due to the error of $v_r$ is drawn \emph{after} the last condition is imposed.  

In the right column of Figure~\ref{fig:fig2}, we plot the relative change in relative velocity by the perspective effects. In situations where the sky-projected separation $s$ is smaller than 30 kau, significant shifts in relative velocities are rare. On the other hand, substantial changes occur frequently when the sky-projected separation $s$ exceeds 30 kau. This is expected because the angular separation tends to be larger, and the relative velocity gets smaller when the sky-projected separation is larger. Note also that the histogram is slightly skewed to the left, which decreases $v_p$ in the largest sky-projected separation.

\begin{figure*}[!htb]
    \centering
    \includegraphics[width=2.0\columnwidth]{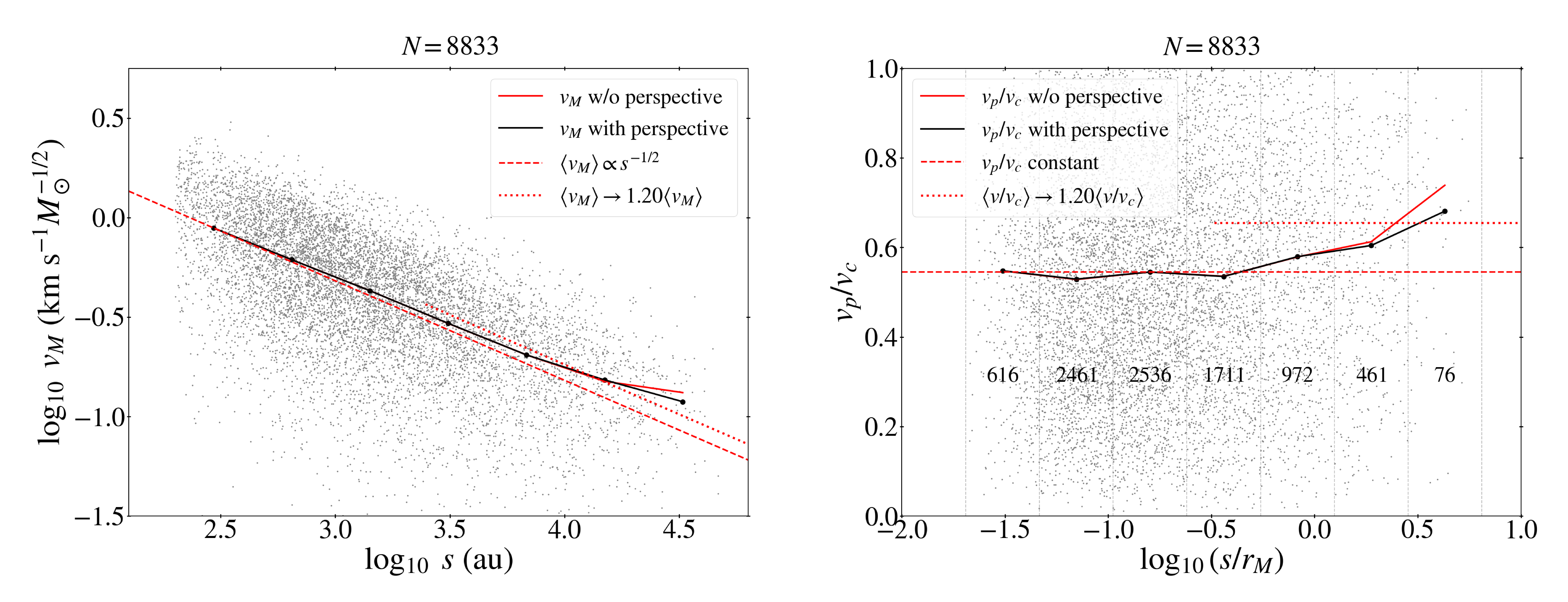}
    \caption{
    \textbf{Left}: Scaling relation for $v_M\equiv v_p/\sqrt{M_{\mathrm{tot}}}$ and $s$. For larger separation, $v_M$ deviates more from the Newtonian scaling relation $v_M\propto s^{-1/2}$. \textbf{Right}: Scaling relation for $\tilde v\equiv v/v_c$ and $s/r_M$. For Newtonian gravity, $\tilde v$ is expected to remain roughly constant if the dependence of eccentricity on $s$ is not considered.}
    \label{fig:scaling}
\end{figure*}

In Figure~\ref{fig:perspec}, we plot the perspective effects as a function of three different variables. We consider $s$, $v_{\rm p(pers)}^2/s$ and $s/r_M$, where $r_M$ is the so-called MOND radius, defined by 
\begin{equation}
    r_M\equiv \sqrt{\frac{GM_{\rm tot}}{a_0}}.
    \label{eq:mondradius}
\end{equation}
This is the length scale where the MOND effect starts to become important. Also, we considered $v_{\rm p(pers)}^2/s$, because the gravitational acceleration is roughly this order. The solid red lines represent the median in each bin. The dashed red lines denote the 16th and 84th percentiles. We also indicate $N$, the number of binaries in each bin. Then we add and subtract the standard deviation divided by $\sqrt N$ to each solid red line to draw the dashed blue lines. In all these graphs, we see that the (approximately) true 2D relative velocity, corrected for the perspective effects, is smaller than the uncorrected value in a statistical sense. 
However, the perspective effects are not significant except for the last bin. In the first graph, it is for $s>$ 30 kau, in the second graph, it is $v_{\rm p(pers)}^2/s < 10^{-11}$ m\,s${^{-2}}$, in the third graph it is $s/r_M>2$. In the last bins, we see that the corrected 2D relative velocities are, on average (more precisely, in the median sense) several percent smaller than the uncorrected ones, 
but not as much as 10\%. In addition, the graphs are skewed ``downward.''

\section{Scaling relations}\label{sec:scalingrelation}
Newtonian gravity predicts the scaling relation $v=\sqrt{GM_{\mathrm{tot}}/r}$ for circular orbits, where $v$ is the norm of the relative 3D velocity and $r$ is the separation between the pair. For realistic elliptical orbits viewed at random orientations, we expect $v_p \propto \sqrt{M_{\mathrm{tot}}/s}$, i.e., $v_M \equiv v_p/\sqrt{M_{\mathrm{tot}}} \propto s^{-1/2}$ in an average or median sense, if we assume that eccentricity is nearly independent of separation for the wide binary sample. Before carrying out MC simulations with realistic eccentricities in the following section, here we examine some scaling relations as a first-order analysis. 

The left panel of Figure~\ref{fig:scaling} exhibits the functional behavior of $v_M(s)$ both with and without the perspective effects.  
We see the clear trend that the median velocity deviates gradually from the Newtonian prediction as $s$ gets larger.
We also see that the perspective effects are only significant at the last bin, the largest separation. This is clearly expected from the origin of the perspective effects. Note also that the perspective effects decrease $v_M$, which agrees with the results of the last section.

The right panel of Figure~\ref{fig:scaling} shows another scaling relation. First, as is common in the literature, we define
\begin{equation}
v_c\equiv\sqrt{\frac{GM_{\mathrm{tot}}}{s}},\quad \tilde v\equiv\frac{v_p}{v_c}.
\label{eq:vtilde}
\end{equation}
Then $s/r_M$ (where $r_M$ is given by Equation~(\ref{eq:mondradius})) and $\tilde v$ are dimensionless. According to Newtonian gravity, the mean (or the median) value of $\tilde v$ should be roughly constant, independent of $s/r_M$ (as long as eccentricity is nearly independent of $s$). However, we clearly see that $\tilde v$ deviates from a flat line beyond about $s/r_M=0.3$.

When eccentricity is taken into account in the analysis, the gravitational acceleration boost would be slightly more prominent than the increase of $\tilde v$ would suggest, considering that a larger separation tends to have a higher eccentricity, which, given a fixed separation $s$, slightly decreases the Newton-predicted velocity. In the next section, we will consider this effect by performing MC simulations. 

\begin{figure}[!htb]
    \centering
    \includegraphics[width=0.95\columnwidth]{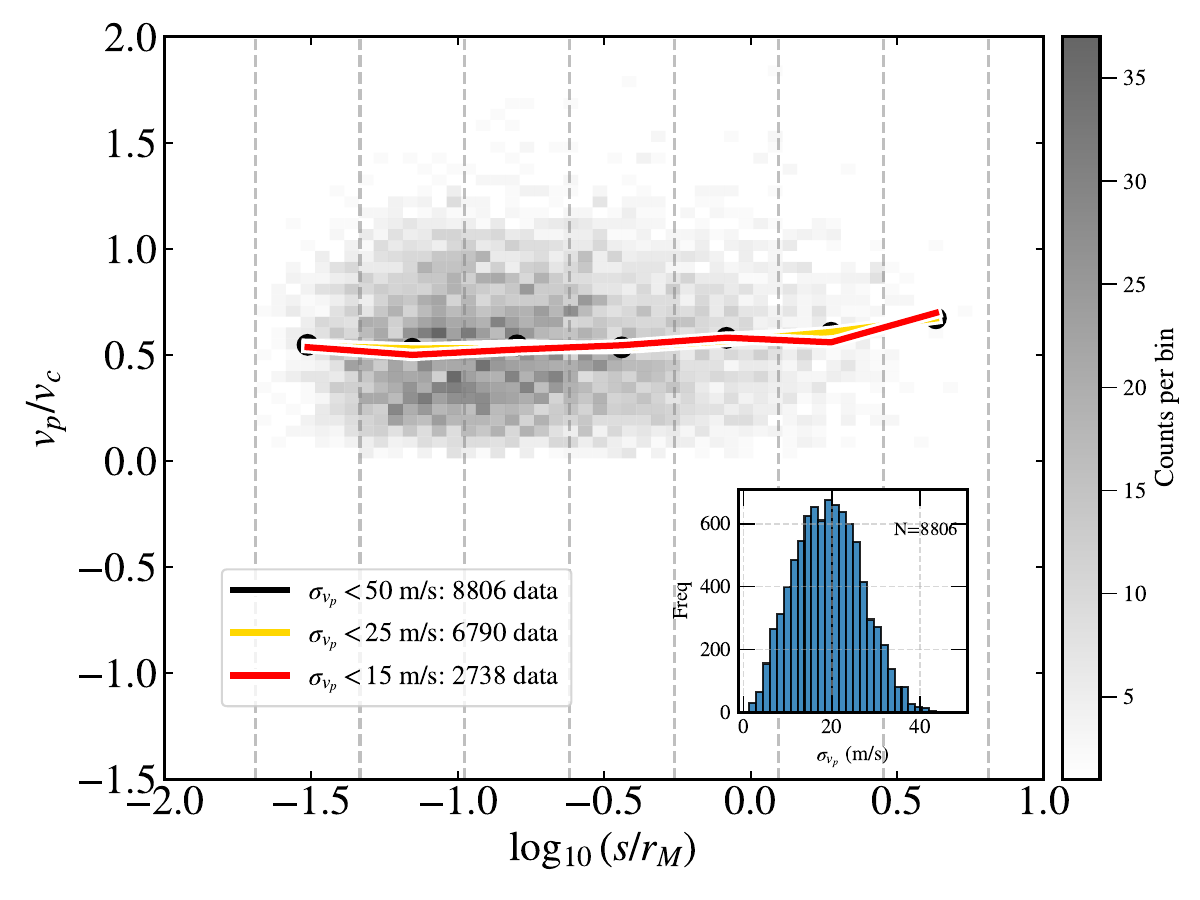}
   \caption{
   The main panel shows the relation between $v_p/v_c$ and $\log_{10}(s/r_M)$ for wide binaries, with running medians indicated for subsamples selected by different projected velocity error ($\sigma_{v_p}$) thresholds. The inset displays the histogram of $\sigma_{v_p}$. The scaling relation is robust to the $\sigma_{v_p}$ selection, with only minor variations due to statistical fluctuations.}
    \label{fig:sigmavp}
\end{figure}

Figure~\ref{fig:sigmavp} further explores the scaling of $\tilde v$ by varying the requirement on the precision of $v_p$. As the inset shows, all $v_p$ for our sample already satisfy $\sigma_{v_p}<~$50 m\,s$^{-1}$, which are quite good. However, we consider subsamples with higher precision as indicated by different colors. We see that there is no tangible dependence on increased precision, apart from the fact that fewer data points increase statistical fluctuations. 

\section{Monte Carlo simulations}\label{sec:MonteCarlo}
In this section, we perform MC simulations to estimate the deviation from Newtonian gravity more accurately by taking into account the dependence of eccentricity on the separation. In particular, we will use two methods from \cite{Chae2023a} and \cite{Chae2024} based on the public codes \citep{Chae2023Zenodo}.

The first method is an orthogonal deviation analysis in an acceleration plane \citep{Chae2023a} (see also Appendix~A of \cite{Chae2023b} for a correction). In this analysis, to be referred to as the ``acceleration-plane analysis'', the following variables are defined: 
\begin{equation}
    g_{\rm N}\equiv \frac{G M_{\rm tot}}{r^2},\quad g\equiv \frac{v^2}{r}\label{eq:gNg},
\end{equation}
where $r$ and $v$ refer to quantities in the 3D space. The quantity $g_{\rm N}$ is the benchmark Newtonian acceleration, while $g$ (referred to as ``kinematic acceleration'' by \cite{Chae2023a}) depends on the phase in the elliptical orbit even for the Newtonian case. Because these quantities are not directly available for the binaries analyzed here, we de-project the observed 2D quantities $s$ and $v_p$ into the 3D space through the \cite{Chae2023a} MC method, considering all statistical possibilities of parameters, including individual eccentricities from \cite{Hwang}.

With MC-generated $g_{\rm N}$ and $g$ at hands, the following are defined:
\begin{equation}
    x\equiv \log_{10} g_{\rm N},\quad y\equiv \log_{10} g,
\end{equation}
from which the orthogonal deviation is defined by the following formula:
\begin{equation}
    \Delta_{\perp}=\frac{y-x}{\sqrt 2},
\end{equation}
which represents a logarithmic deviation from the Newtonian circular prediction of a given model. Note that $\Delta_{\perp}$ is calculated for both the observational data and the Newtonian simulation data as $\Delta_{\perp}\neq 0$ for Newtonian elliptical orbits. For each data point $(x,y)$, \cite{Chae2023a} defines a variable $x_0\equiv x+(y-x)/2$ so that $(x_0,x_0)$ corresponds to the orthogonal projection of $(x,y)$ onto the diagonal line. Then, bins of $x_0$ are defined, and statistical properties of $\Delta_{\perp}$ are calculated for each bin. In particular, the binned medians $\langle\Delta_{\perp}\rangle_{\rm{obs}}$ and $\langle\Delta_{\perp}\rangle_{\rm{newt}}$ are obtained for the observational and Newtonian simulation data. Finally, the relative orthogonal deviation, i.e., the difference between the two is defined as
\begin{equation}
    \delta_{\rm obs-newt} \equiv \langle\Delta_{\perp}\rangle_{\rm{obs}} - \langle\Delta_{\perp}\rangle_{\rm{newt}}.
    \label{eq:deviation}
\end{equation}
The value and its uncertainty of $\delta_{\rm obs-newt}$ are estimated from a number of MC samples. Then, the gravity boost factor, i.e.\ the ratio of the observational and Newton-predicted kinematic accelerations, is derived by 
\begin{equation}
    \gamma_g =g_{\rm obs}/g_{\rm pred}=10^{\sqrt 2 \delta_{\rm obs-newt}}.
    \label{eq:gobsgpred}
\end{equation}

\begin{figure*}
    \centering
    \includegraphics[width=2.0\columnwidth]{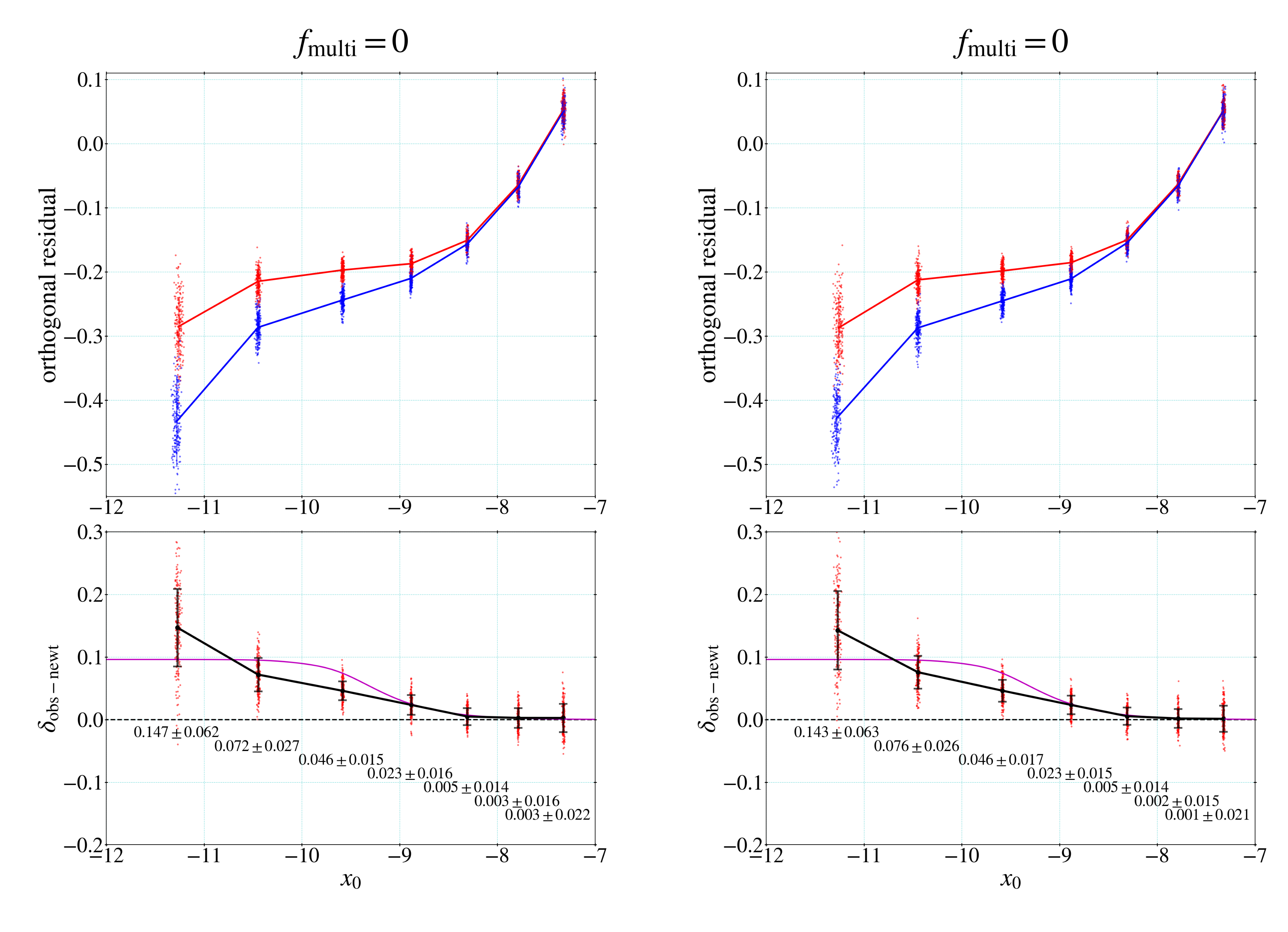}
   \caption{\textbf{Left}: The observed acceleration and the Newtonian acceleration obtained by MC simulation, using the method of the acceleration-plane analysis by \cite{Chae2023a}. See the main text for the explanation. \textbf{Right}: Same as left, but with Gaia FLAME masses for 42.7\% of stars. There is no tangible difference between the two figures.}
    \label{fig:simulationacceleration}
\end{figure*}

In Figure~\ref{fig:simulationacceleration}, we present our results from the acceleration-plane analysis. The left column represents the result based on the nominal values of stellar masses following \cite{Chae2023a}, while the right column represents an alternative result based on Gaia FLAME masses (see below). Unless stated otherwise, any quoted values in the following are from the nominal result.

The upper panel shows the values of $\langle\Delta_{\perp}\rangle_{\rm{obs}}$ and $\langle\Delta_{\perp}\rangle_{\rm{newt}}$ for seven bins of $x_0$ from 200 MC simulations.
Their difference, $\delta_{\rm obs-newt}$ (Equation~(\ref{eq:deviation})), is shown in the bottom panel as a function of $x_0$. 
In the figure, $f_{\rm multi}$ denotes the fraction of binaries that have hidden (faint) star(s) not distinguished or resolved from the two stars by the Gaia data (recall that all resolved multiple systems have already been excluded from our sample). 
Such hidden stars may have biased the observed relative velocity between the pair and the estimate of the total mass based on their luminosities,  
and thus need to be taken into account. Usually, $f_{\rm multi}(>0)$ is tuned to match the observed accelerations in the high acceleration range, where there should be no deviation from Newtonian predictions. 

For the present sample, the two highest acceleration bins with $x_0 > -8$ satisfy $\delta_{\rm obs-newt}\approx 0$ with $f_{\rm multi}=0$. This means that in an average or median sense, our sample is pure, free of kinematic contaminants. This can be understood from the fact that every binary in our sample satisfies the highest data qualities (e.g., $\sigma_{v_p}<50$~m\,s$^{-1}$) and stringent selection criteria such as {\tt ruwe} $< 1.2$. This will not mean that there are no individual exceptions that evaded our selection criteria, but relatively few exceptions would not affect the median trend. We also emphasize that every star from all our binaries satisfies the same criteria with an equal probability, regardless of their membership, meaning that the exceptional cases would occur regardless of the binary separation and thus cannot bias the median trend we see. Although a close binary of the same mass and separation from the host star can have a relatively larger impact on the relative velocity in the wide binary of a larger separation, it would not affect the median trend significantly as long as its probability is low because a median is immune to a few exceptional outliers.

The purple line in Figure \ref{fig:simulationacceleration} is the AQUAL prediction, as presented in Figure~1 of \cite{Chae2023b}. We find that the gravitational acceleration boost factor $\gamma_g$ (Equation~(\ref{eq:gobsgpred})) is $1.61^{+0.37}_{-0.29}$ in the last bin, i.e., at $g_{\rm N}\approx 10^{-11.0}\text{m s}^{-2}$ and $1.26^{+0.12}_{-0.10}$ in the second to last bin, i.e., at $g_{\rm N}\approx 10^{-10.3}\text{m s}^{-2}$. The last bin represents the lowest-acceleration bin probed to date, thanks to our inclusion of wide binaries with $s>30$~kau. Unfortunately, it returns a result with a relatively large uncertainty. Nevertheless, it is intriguing to note that the boost factor $1.61^{+0.37}_{-0.29}$ is consistent with the AQUAL/QUMOND prediction $\approx 1.4-1.5$. Although not shown in the figure, we find $1.32^{+0.12}_{-0.11}$ for $g_{\rm N}\la 10^{-10}\text{m s}^{-2}$ that encompasses the two bins. These boost factors are consistent with previous estimates based on differently defined samples (e.g., \citealt{Chae2023a,Chae2023b,Chae2024,Chae2025,Hernandez:2023upr,HernandezKroupa}).

\begin{figure*}
    \centering
    \includegraphics[width=0.9\linewidth]{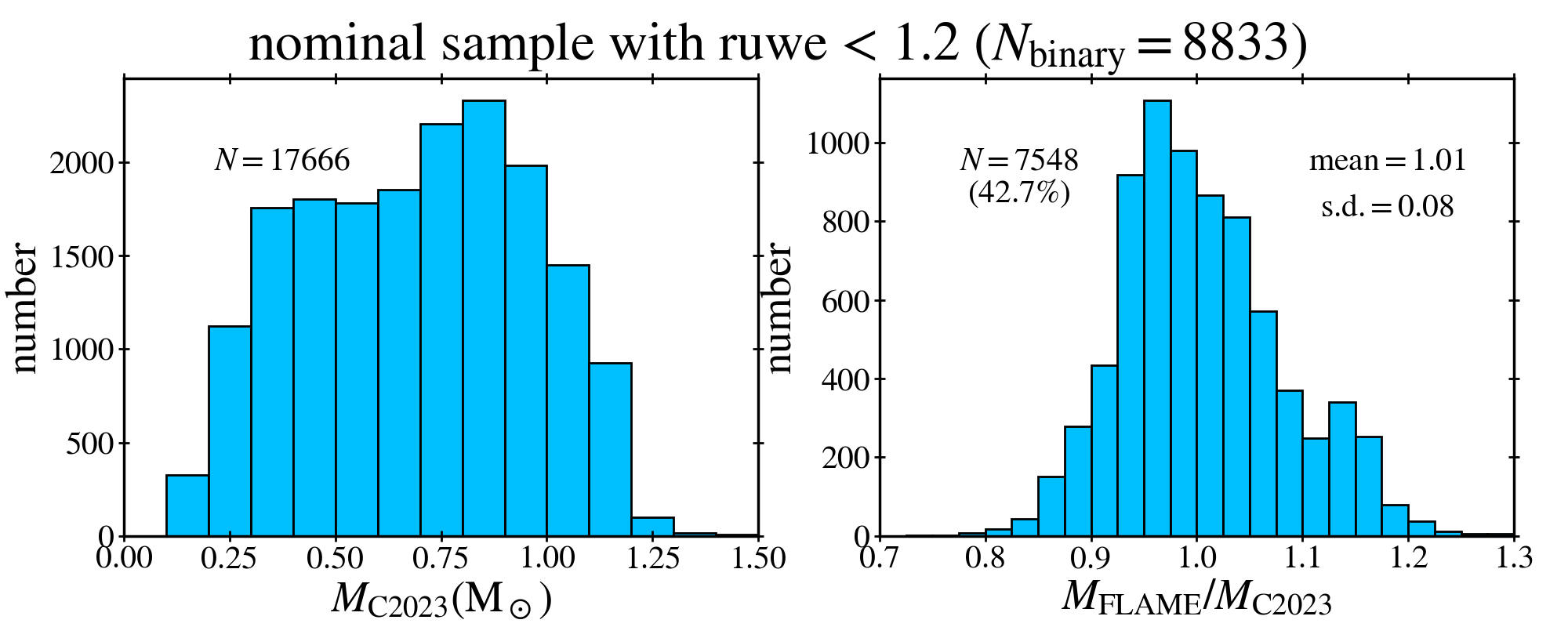}
    \caption{\textbf{Left}: Distribution of stellar masses obtained by the method of \cite{Chae2023a} for the nominal sample used in Figure~\ref{fig:simulationacceleration}. \textbf{Right}: Comparison between the Gaia FLAME masses and the \citep{Chae2023a} masses. They agree well on average, but there is about 8\% scatter.}
    \label{fig:flamemass}
\end{figure*}

The alternative result shown in the right column of Figure~\ref{fig:simulationacceleration} is almost indistinguishable from the nominal result. 
This is not surprising as the nominal and FLAME masses of stars (whenever the latter are available) agree well with each other statistically as shown in Figure~\ref{fig:flamemass}. Although there is $\approx 8$\% scatter for the ratio of the two masses, the mean/median is close to 1.

As a second statistical method, we consider an analysis of the normalized 2D velocity $\tilde v (\equiv v_p/v_c)$ (Equation~(\ref{eq:vtilde})) without a deprojection to the 3D space. This method is considered in part because the parameter ${\tilde v}$ has been popular in the literature of wide binary stars (e.g., \citealt{Banik:2018,Pittordis:2018,Pittordis:2019}). We use the approach and code presented in \cite{Chae2024} that are similar in spirit to those of the acceleration-plane analysis, except that no deprojection is considered. In this approach, gravity is probed by the statistical properties of $\tilde v$ in bins of $s/r_{\rm M}$. These parameters serve as proxies for $g$ and $g_{\rm N}$ (Equation~(\ref{eq:gNg}): see Section~2.2 of \cite{Chae2024}). Parameters $\tilde v$ and $s/r_{\rm M}$ are more directly accessible from observations than $g$ and $g_{\rm N}$, but the latter probe gravity more directly and accurately. For the observed sample, many MC realizations of Newtonian gravity are produced, and the distributions of the binned medians from the MC samples are compared with the observed medians. Each binary in a mock Newtonian sample has the same projected separation $s$ and the same total mass $M_{\rm tot}$ as the observed binary, but differs only in the projected velocity $v_p$: the mock Newtonian binary has a value of $v_p$ randomly realized with the observed property of the binary assuming Newtonian gravity.

\begin{figure}[!htb]
    \centering
    \includegraphics[width=0.95\linewidth]{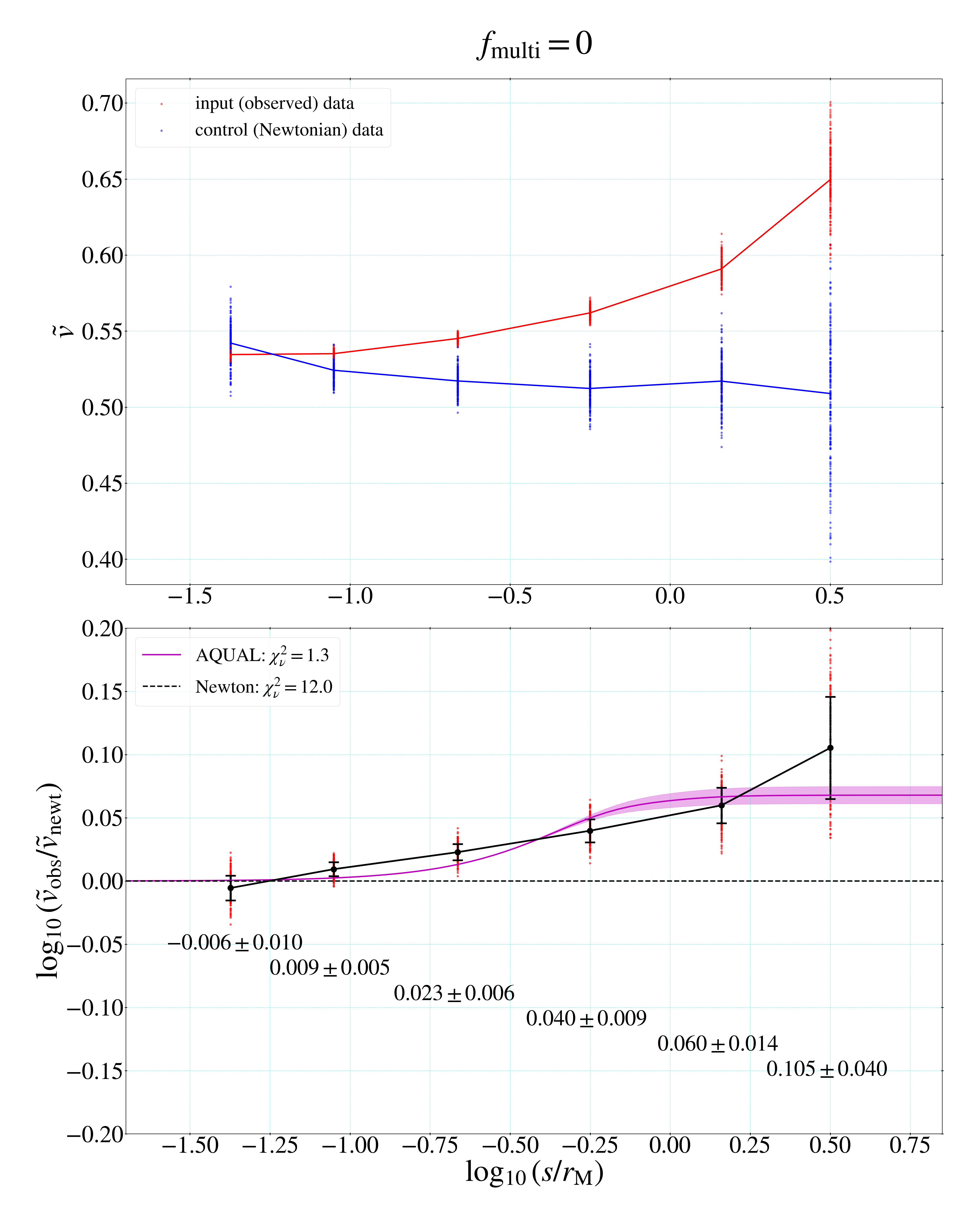}
    \caption{MC simulation estimate of the Newtonian $\tilde v$ and its comparison with the observed value. Newtonian gravity is excluded, while there is a good agreement with AQUAL.}
    \label{fig:vtilde}
\end{figure}

We present our simulation results in Figure~\ref{fig:vtilde}. We find that Newton-simulated $\tilde v$ \emph{decreases} mildly for our sample as the separation (or more precisely, $s/r_M$) gets larger. In an ideal sample where the binaries in each bin have statistically similar eccentricities and are free of biases due to artificially imposed certain selection criteria, Newton-predicted $\tilde v$ is expected to be flat. However, a real sample, apparently as in our case, can exhibit a non-flat Newtonian behavior due to varied eccentricities and/or sample selection criteria. Due to this possibility, it is important to calculate the sample-specific Newtonian benchmark against which the observed $\tilde v$ should be compared. In other words, assuming that the Newtonian prediction is automatically flat without calculating it specifically for one's sample, he/she can have a wrong conclusion about whether the observed $\tilde v$ is deviating from Newton and how much if so. This is because both the observed and the Newtonian $\tilde v$ will suffer from similar effects due to varied eccentricities or artificial selection criteria (see, e.g., Figure~8 of \cite{Chae2024} for the latter).  

\begin{figure}[!htb]
    \centering
    \includegraphics[width=0.95\linewidth]{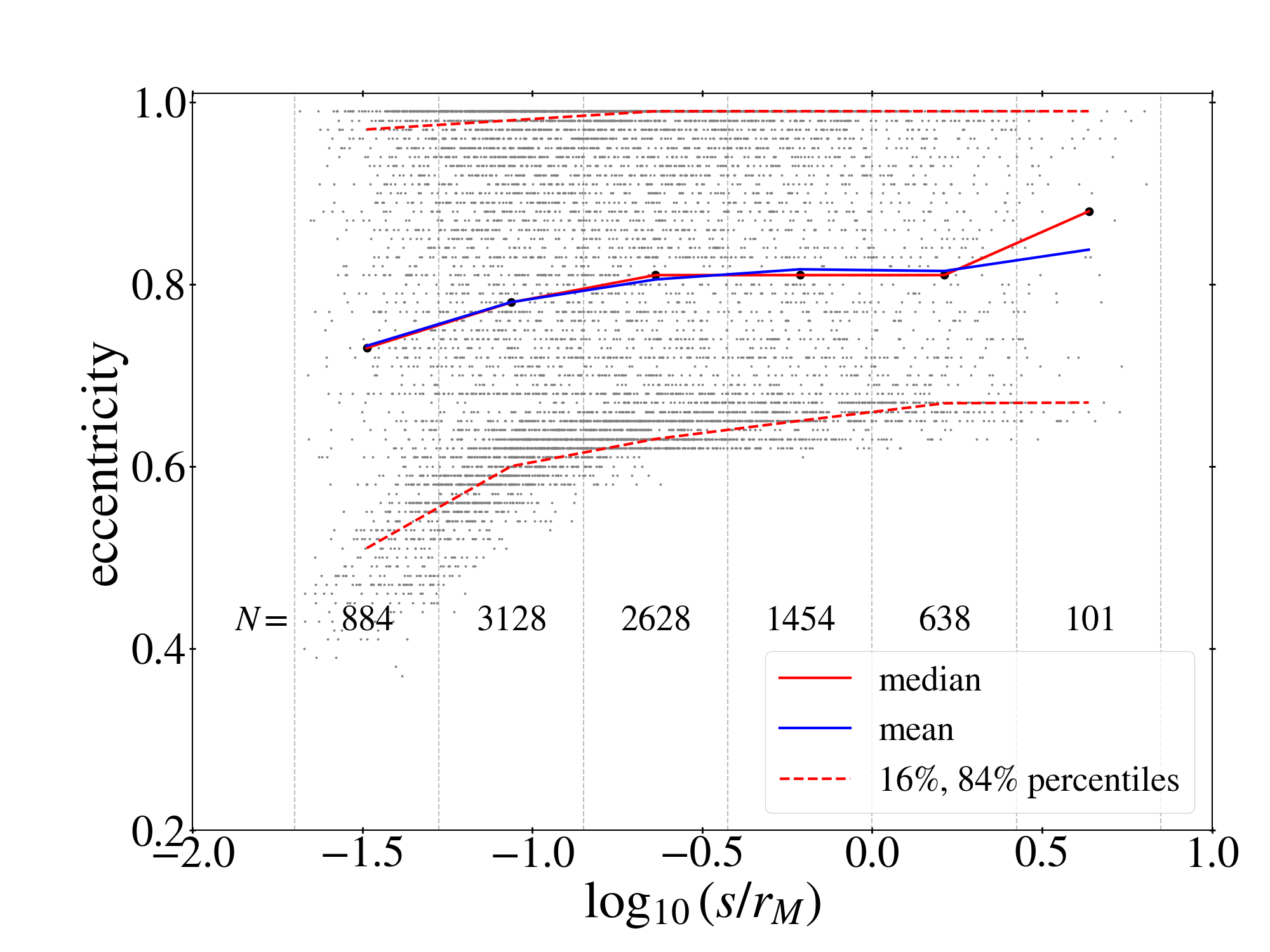}
    \caption{Eccentricity medians and means for the bins in Figure~\ref{fig:vtilde}. The trend in the rise of eccentricity with $s/r_{\rm M}$ explains the trend of decline in the Newtonian acceleration in that figure. 
}\label{fig:eccentricity}
\end{figure}

We note that in the present case, for our sample, the mild decline of the Newtonian prediction is due to the increase of eccentricity for larger separation $s$. Even though this tendency is mitigated when we change our variable from $s$ to $s/r_M$, its effect is still present. See Figure~\ref{fig:eccentricity} for the scaling of eccentricity with $s/r_M$. The eccentricity medians for the first bin and the second bin are 0.73 and 0.78, but the ones for the other bins are 0.81, 0.81, 0.81, 0.88. Regarding the other bins, the means of the third, fourth, fifth, and sixth bins are 0.805, 0.816, 0.814, and 0.838, showing a more or less increasing tendency. This agrees with the relatively sharp decline of $\tilde v$ in the first two bins, followed by only a mild decline in the consecutive bins, and the relatively steep decline again in the last bin. Thus, properly calculating the Newtonian prediction of ${\tilde v}$ for our sample by accounting for the dependence of eccentricity on separation not only fails to explain the deviation from Newtonian gravity, but actually exacerbates it. If the MOND effects in wide binaries were false signals due to the eccentricity dependence on separation, wide binaries with larger separations would have to have smaller eccentricities, but the trend is the direct opposite.

We also see that AQUAL is strongly favored with $\chi_\nu^2=1.3$, while Newtonian gravity is strongly ruled out with $\chi_\nu^2=12.0$. Here we have obtained the $\chi_\nu^2$ (i.e., reduced $\chi^2$) values based on Equation~(21) of \cite{Chae2024}. 

\begin{figure}[!htb]
    \centering
    \includegraphics[width=0.95\linewidth]{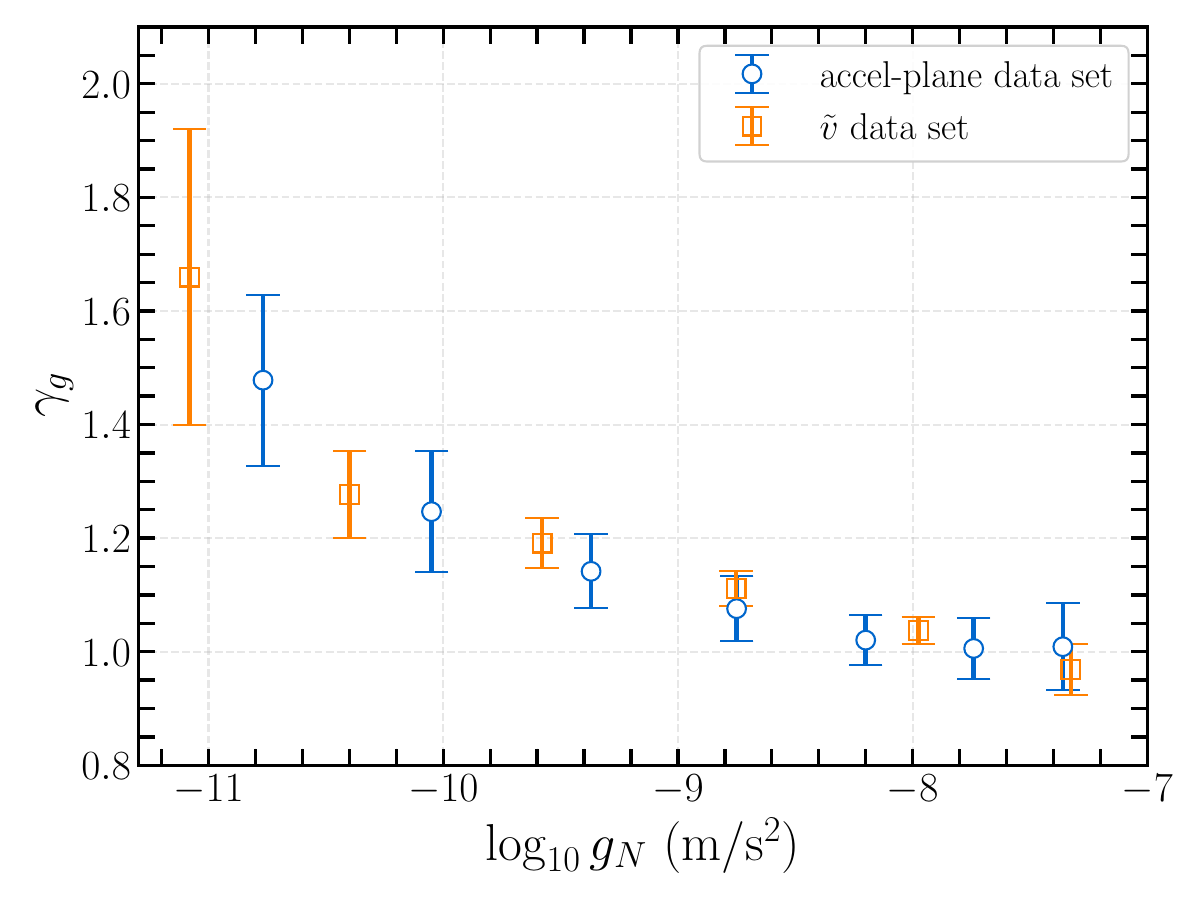}
    \caption{
    Comparison of $\gamma_g$ as a function of Newtonian acceleration $g_{\rm N}$, derived from the orthogonal residual in the acceleration plane (blue circles) and $\tilde v$ (orange squares) methods. The results from both approaches are in excellent agreement across the full range of $g_{\rm N}$, demonstrating the robustness of the measurement technique.
    }
    \label{fig:comparison}
\end{figure}

In Figure~\ref{fig:comparison}, we summarize our results from the two methods by plotting values of $\gamma_{g}$ (gravitational acceleration boost factor) as a functions of Newtonian acceleration $g_{\rm N}$ (Equation~(\ref{eq:gNg})). For the first method, $\gamma_{g}$ is given by Equation~(\ref{eq:gobsgpred}). For the second method, we estimate it by 
\begin{equation}
    \gamma_g=(\tilde v_{\rm obs}/\tilde v_{\rm newt})^2,
    \label{eq:gammag_vtilde}
\end{equation}
and $g_{\rm N}$ is estimated by Equation~(23) of \cite{Chae2024}.
We see that the two sets of results agree with each other overall. The nominal error bars from the ${\tilde v}$ analysis are somewhat smaller, but this does not mean that it is more accurate because it involves statistical approximations in estimating both $\gamma_g$ and $g_{\rm N}$, while the acceleration-plane analysis takes into account all statistical possibilities of the parameter space without approximations. In particular, the acceleration-plane analysis is known to reproduce well the flat Newtonian behavior (see, e.g., \cite{Chae2024}) at high acceleration $g_{\rm N}\ga 10^{-8}$~m\,s$^{-2}$ as is also evident from our results. Nevertheless, the good overall agreement indicates that the ${\tilde v}$ analysis (based on the proper Newtonian prediction) is appropriate at least as a first-order approximation. 

The above results are based on the nominal sample and the 2D velocities corrected for the perspective effects.
In Appendix \ref{sec:appendixsansperspective}, we perform the acceleration-plane analysis of our nominal sample without considering the perspective effects, so that the impact of the perspective effects on gravity inference can be seen directly. We see that the perspective effects have a rather minor, though non-negligible, impact on gravity inference. In Appendix \ref{sec:appendixruwe}, we consider alternative samples with varied {\tt ruwe} limits of $<1.1$ and $<1.3$ instead of the nominal limit $<1.2$. We see that there are no significant changes in the gravity inference. In Appendix \ref{sec:belokurov}, we consider a sample obtained by a stricter criterion for the color-magnitude (CM) relation than the one of \cite{Penoyre}. We mask the region in the CM diagram that may contain the photometric binary sequence (note however that some in the masked region may well be single stars). It turns out that this cut has a minor impact on gravity inference because those removed are relatively few and therefore cannot affect the median trend significantly. 
In Appendix \ref{sec:bothRV}, we consider the sample of binaries with both radial velocities available. 
Compared to the nominal sample, gravity inference is little changed.

\section{Discussion and Conclusion}\label{sec:conclusion}

In this study, we have considered very widely (projected separation $s>30$~kau) separated binaries as an attempt to probe gravity at extremely low internal acceleration $\la 10^{-11}$~m\,s$^{-2}$ that is an order-of-magnitude lower than the MOND critical acceleration $a_0$. Because gravitationally-bound isolated binaries are increasingly rare and more difficult to identify as $s$ gets larger, we have searched the Gaia archive up to a distance of 300~pc and a separation of 50~kau. This is chosen as a compromise between the search volume size and the required data qualities, as data qualities become increasingly poorer as distance increases.  Our search limit is larger than in previous studies by Chae \citep{Chae2023a,Chae2023b,Chae2024} and Hernandez et al.\ \citep{Hernandez:2023qfj,Hernandez:2023upr} who considered only up to a distance of 200~pc and separation of 30~kau.

Although our sample is limited by the relatively small number of isolated wide binaries with $s>30$~kau, it is a new sample including such wide binaries selected with very stringent selection criteria so that kinematic contaminants are minimized. We have explicitly corrected the observed PMs for the perspective effects. We then applied two statistical methods to probe the median trend in bins of acceleration or normalized separation. Because our methods rely on the medians in the bins, any small fraction of kinematic contaminants would not matter for our study.

As regards any possible bias due to unremoved kinematic contaminants, we note two facts. First, every single star from our sample satisfies the same data qualities, such as parallax and PM fractional errors and Gaia ruwe value, regardless of the property of the hosting binary. Because kinematic contaminants such as hidden close companion stars or massive Jovian planets occur on a single-star basis, our uniform selection means that all bins of binaries will suffer from similar probabilities of contamination, and thus the median trend is not expected to be affected significantly by any small amounts of hidden contaminants. Second, at highest-acceleration or smallest-separation bins, the observed median accelerations automatically match the Newtonian predictions without including any kinematic contaminants, indicating that exceptionally survived kinematically contaminated cases are rare overall.

Our new results based on the sample including very wide binaries with the perspective effects, confirm the previous finding that the observed gravitational acceleration is larger than the Newtonian prediction in low accelerations. Thus, the low-acceleration gravitational anomaly in wide binaries cannot be explained away by the perspective effects. Eccentricity variation with separation cannot remove the anomaly either. To see why, it is helpful to consider the functional behavior of $\tilde v (s/r_{\rm M})$. The observed trend from our sample, represented by the red curve in Figure~\ref{fig:vtilde} increases with $s/r_{\rm M}$, while the naive/default Newtonian prediction is flat. However, if we take into account eccentricities in the bins to correctly calculate the Newtonian prediction of $\tilde v (s/r_{\rm M})$, it actually declines mildly with $s/r_{\rm M}$ as shown by the blue curve. Thus, taking into account the dependence of observationally-inferred eccentricity on separation in our analysis increases the anomaly (compared with the naive case without calculating the Newtonian prediction), rather than decreasing it, let alone eliminating it. In fact, the comparison between the observed curve and the Newtonian curve restores the correct value of $\gamma_g$ obtained by the acceleration-plane analysis.  

In the lowest-acceleration bin with $g_{\rm N}\la 10^{-11}$~m\,s$^{-2}$, we find that gravitational boost is significant. While the value of $\gamma_g=1.61_{-0.29}^{+0.37}$ (based on the acceleration-plane analysis) from this bin alone is more than $2\sigma$ away from Newtonian gravity ($\gamma_g = 1$), it is well consistent with $\gamma_g \approx 1.4$ predicted by classical Lagrangian models \citep{Bekenstein:1984tv,Milgrom:2009ee} of MOND gravity. This agreement, however, should be taken with a grain of salt because a proper comparison of the wide binary data with those MOND models requires more realistic numerical simulations than considered in the literature (e.g., \citealt{Chae:2022rdr}).

Our study indicates that the probe of gravity in the low-acceleration limit with $s>30$~kau will be quite limited even in the future, when based only on conventional statistical methods as considered here. This is because the number of isolated wide binaries in the solar neighborhood may not be sufficient to obtain a desired precision of $\gamma_g$. This difficulty may be overcome by the Bayesian 3D methodology \citep{Chae2025,Chae2025b} with accurate measured radial velocities in the future, because it can return a precise determination of $\gamma_g$ even with dozens of wide binaries once high-precision radial velocities become available. 

Nevertheless, an issue still remains. Our maximum 2D separation, 50 kau, corresponds to 0.24 pc, which is far below 1.7 pc, the tidal radius of a binary with the total mass of 2$M_{\odot}$ \citep{Jiang}. However, 3D separation can be easily two or three times the 2D separation, and considering the high eccentricity of wide binaries with very large separations, their separation can be large enough to exceed the tidal radius, during the \emph{course} of the orbit, even though they may not be at the moment of the observation. In fact, we estimate that about 15\% of our observed binaries with $s>30$~kau exceed the tidal radius during their orbits.  
In this calculation, we considered the semi-Newtonian approximation, i.e., replacing $M_{\rm tot}$ by $\gamma_g M_{\rm tot}$ to obtain the tidal radius. 
If a binary is expected to exceed the tidal radius during its orbit, it means that it is likely to be a fly-by rather than a gravitationally-bound system. This judgment cannot be made for a particular individual system based only on the 2D data. A reconstruction of orbits with 3D data is needed as suggested recently \citep{Chae2025b}. However, we note that the sample used in this work is likely to contain a relatively minor fraction and all our selected binaries already have velocity bounds (Equation~(\ref{eq:velbound})), and therefore the median statistic is likely to be not significantly affected.

\begin{acknowledgments}
We thank the reviewer for the constructive feedback and valuable suggestions.
This work was supported by the National Research Foundation of Korea (NRF-2022R1A2C1092306).
YT also acknowledges Taiwan National Science and Technology Council (NSTC) grant 114-2112-M-008-024-MY3.
\end{acknowledgments}

\bibliographystyle{aasjournalv7.bst}
\bibliography{main.bib}{}

\appendix
\twocolumngrid

\section{Acceleration-plane analysis without considering the perspective effects}\label{sec:appendixsansperspective}
In this appendix, we perform the acceleration-plane analysis for our nominal sample without considering the perspective effects, to compare it with Figure~\ref{fig:simulationacceleration}. We find that the two lowest acceleration bins have slightly higher accelerations than the ones for which the perspective effects are taken into account. $\delta_{\rm obs-newt}$ is 0.160 for the lowest bin, but it is 0.147 with the perspective effects. For the second-lowest bin, it is 0.079 compared to the corrected value, 0.072. This agrees with our earlier results that correctly taking into account the perspective effects slightly decreases the deviation from Newtonian gravity, but only for very low accelerations, where the deviation is already large.

\begin{figure}[!htb]
\centering
\includegraphics[width=0.9\columnwidth]{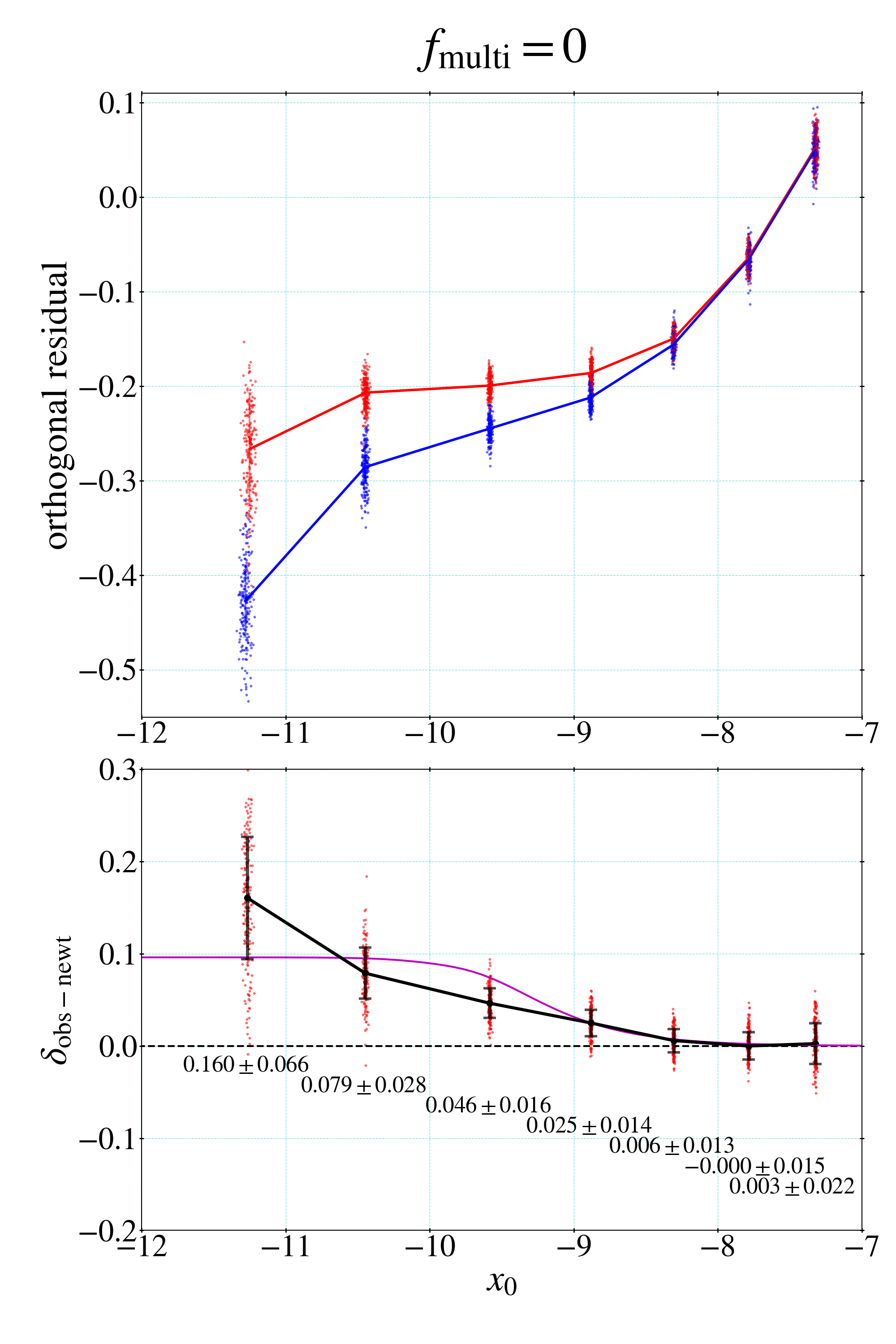}
\caption{Acceleration-plane analysis of our nominal sample without considering the perspective effects. Compared with Figure~\ref{fig:simulationacceleration}, the two lowest acceleration bins have slightly higher $\delta_{\rm obs-newt}$ values.
}\label{fig:sansperspective}
\end{figure}

\section{Results with varied {\tt ruwe} limits}\label{sec:appendixruwe}
\begin{figure*}[!htb]
\centering
\includegraphics[width=2.0\columnwidth]{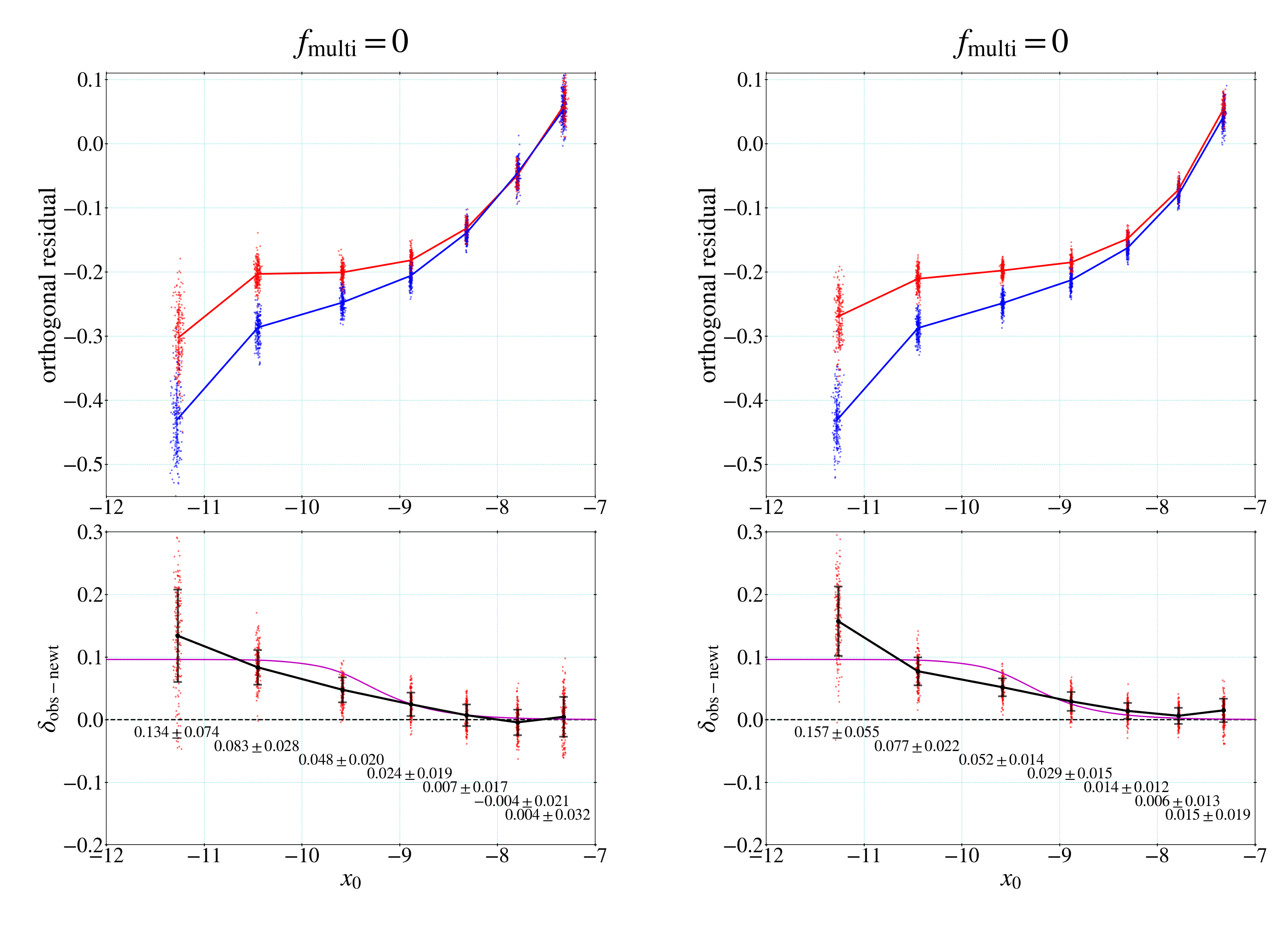}
\caption{
\textbf{Left}: Same as Figure~\ref{fig:simulationacceleration}, but with ${\tt ruwe}<1.1$. \textbf{Right}: Same as Figure~\ref{fig:simulationacceleration}, but with ${\tt ruwe}<1.3$. These results with variations of ${\tt ruwe}$ limit indicate that the nominal results are robust against the ${\tt ruwe}$ variation within the considered range. 
}\label{fig:ruwe}
\end{figure*}

In this appendix, we consider different ${\tt ruwe}$ limits for Figure~\ref{fig:simulationacceleration}, which is based on ${\tt ruwe}<1.2$. For fair comparisons, we take $f_{\rm multi}=0$ for all cases. In Figure~\ref{fig:ruwe}, we see that the case with ${\tt ruwe}<1.1$ is almost indistinguishable from the nominal case with ${\tt ruwe}<1.2$. However, the case with ${\tt ruwe}<1.3$ hints at a minor tendency for the observed acceleration to deviate from the Newtonian one at the highest acceleration bins. This may hint that some kinematically contaminated systems have started to enter the sample. Note, however, that the deviation is not statistically significant. Thus, our results based on ${\tt ruwe}<1.2$ are robust against some reasonable variation of the ${\tt ruwe}$ limit. 

\section{Robustness Test against Contamination from Photometric Binaries}\label{sec:belokurov}

In this appendix, we test the robustness of our main results against potential contamination from unresolved close binaries, often referred to as photometric binaries. As noted in studies such as~\cite{Belokurov, Hernandez:2023qfj}, these systems can be identified on a Color-Magnitude Diagram (CMD) as they typically form a sequence located above the single-star main sequence due to photometric blending. While our primary sample selection, which includes a ${\tt ruwe} < 1.2$ cut, is effective at removing many unresolved multiples, some may remain and potentially bias the kinematic analysis.

\begin{figure*}[!htb]
\centering
\includegraphics[width=2.0\columnwidth]{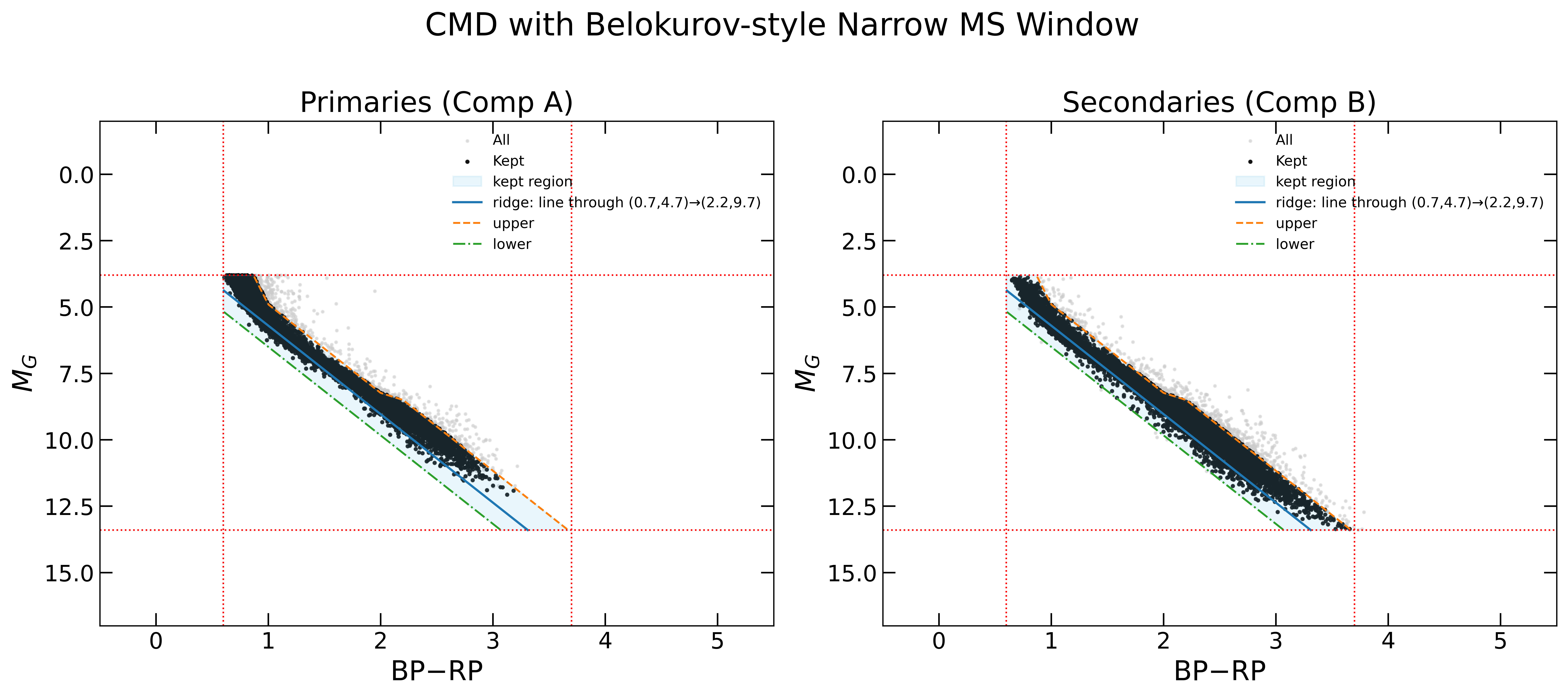}
\caption{
We apply additional cuts in the CMD to exclude the photometric binary sequence lying above the main sequence (Figure~\ref{fig:HRdiagram}), which is produced by unresolved close binaries. After this cleaning, 7,800 binaries remain from the initial 8,833 pairs.
}\label{fig:CMD}
\end{figure*}

To address this, we implement a more conservative CMD selection strategy inspired by the methodology of~\cite{Hernandez:2023qfj}. We define a main-sequence ridge line by connecting the points (BP-RP, $M_G$) = (0.7, 4.7) and (2.2, 9.7). We then construct a selection band around this line, designed to follow the observed main sequence more closely. Specifically, we exclude any binary system where either component lies more than 0.8 magnitudes above or below this ridge line. While the lower boundary is kept at 0.8 magnitudes below, the upper boundary is further adjusted as a function of color to account for the main sequence's intrinsic curvature, ensuring a precise exclusion of the photometric binary population as illustrated in Figure~\ref{fig:CMD}. We note that there can be single stars above the upper boundary and the cut is not designed to selectively remove only photometric binaries. The application of this stricter cut reduces our sample size from 8,833 to 7,800 pairs.

\begin{figure}[!htb]
\centering
\includegraphics[width=0.9\columnwidth]{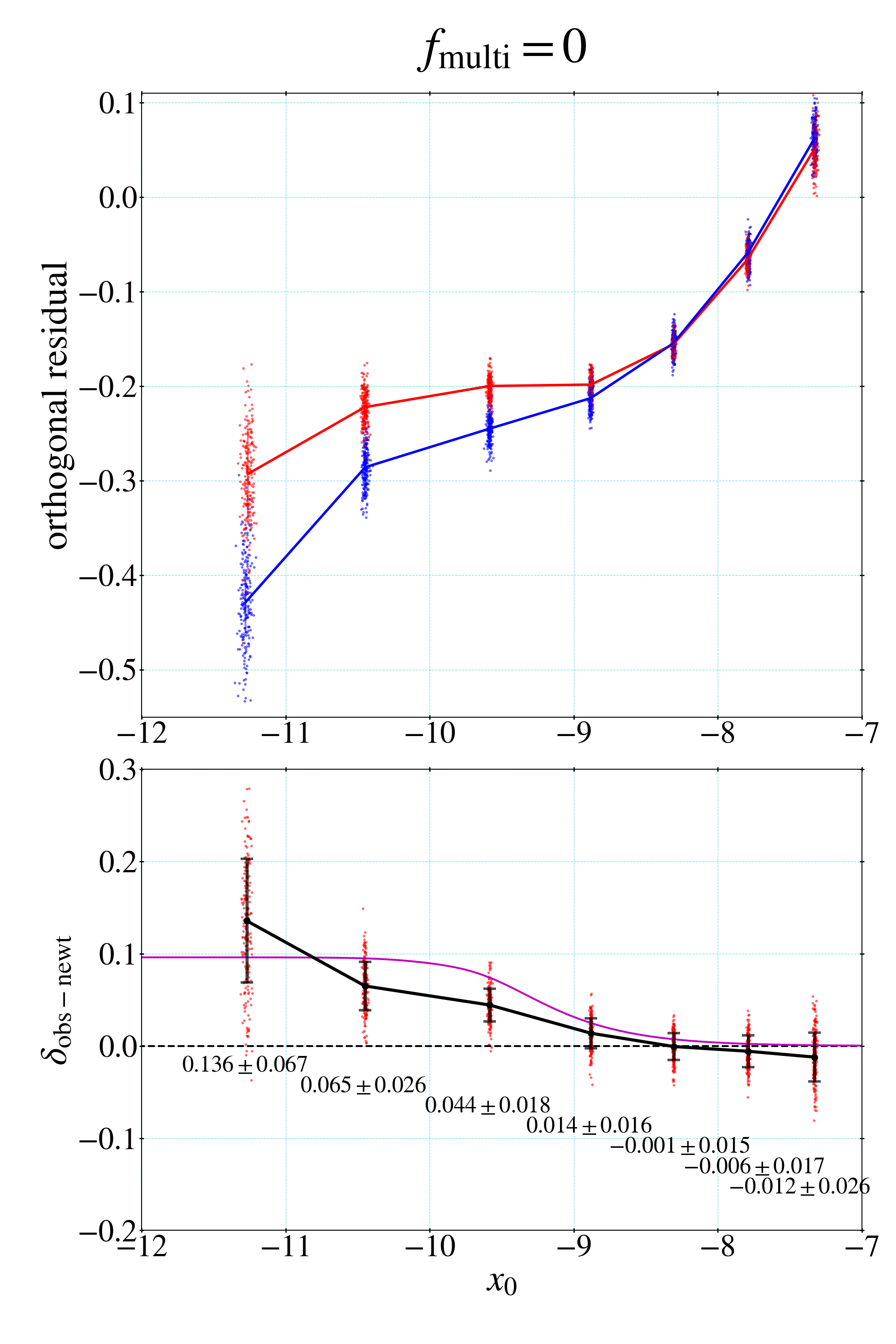}
\caption{Acceleration-plane analysis of our sample obtained by removing the potential contamination from unresolved close binaries as in Figure~\ref{fig:CMD}. Compared with Figure~\ref{fig:simulationacceleration}, there is no 
significant difference.
}\label{fig:belokurov}
\end{figure}

The results of the acceleration-plane analysis using this cleaner sample are presented in Figure~\ref{fig:belokurov}. When compared with the analysis of our nominal sample (Figure~\ref{fig:simulationacceleration}), the magnitude of the observed gravitational anomaly is slightly attenuated, but the results remain quantitatively consistent and the statistical significance of the deviation from the Newtonian prediction is largely unchanged. We therefore conclude that our primary findings are robust, and the observed anomaly cannot be attributed to kinematic contamination from the population of photometric binaries removed in this test.

\section{Results for binaries with both radial velocities available}\label{sec:bothRV}
\begin{figure}[!htb]
\centering
\includegraphics[width=0.9\columnwidth]{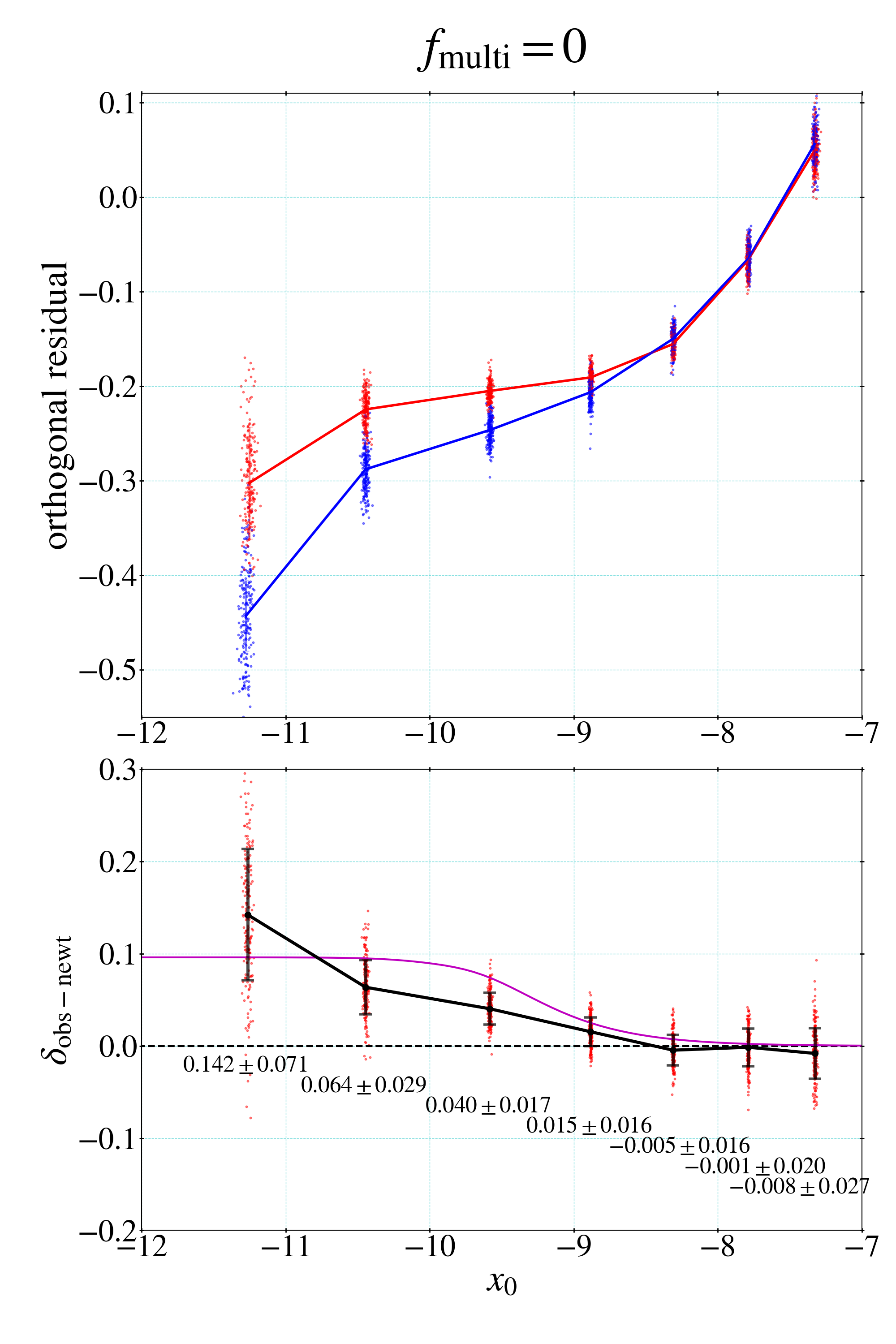}
\caption{Acceleration-plane analysis for binaries with both radial velocities available. The number of binaries is 7,250, slightly smaller than our nominal sample 8,833. There is no appreciable difference from Figure \ref{fig:simulationacceleration}.
}\label{fig:fig2RV}
\end{figure}

As mentioned before, in our analysis so far, we considered binaries with both radial velocities available, or one of the radial velocities available. In this appendix, we present the acceleration-plane analysis for binaries with both radial velocities available. In Figure~\ref{fig:fig2RV}, we see that there is no tangible difference from Figure~\ref{fig:simulationacceleration}.

\end{document}